\documentclass[journal]{IEEEtran}

\usepackage{xcolor,soul,framed}

\usepackage[pdftex]{graphicx}
\graphicspath{{../pdf/}{../jpeg/}}
\DeclareGraphicsExtensions{.pdf,.jpeg,.png}
\usepackage[cmex10]{amsmath}

\usepackage{array}
\usepackage{mdwmath}
\usepackage{mdwtab}
\usepackage{eqparbox}
\usepackage{upgreek}
\usepackage{url}
\usepackage{tabu}
\usepackage{xcolor}
\usepackage{nicefrac}
\usepackage[export]{adjustbox}[2011/08/13]

\newcommand{\ket}[1]{\ensuremath{|#1\mkern-1mu\rangle}}
\newcommand{\ovalap}[2]{\ensuremath{\langle  #1 | #2 \rangle }}
\newcolumntype{P}[1]{>{\centering\arraybackslash}p{#1}}

\definecolor{myblue}{rgb}{0.03, 0.23, 0.63}

\newcommand{\ch}[1]{{\color{black}#1}}

\begin{document}
\title{\ch{Advances in silicon quantum photonics}} 
    
\author{Jeremy C. Adcock, Jueming Bao, Yulin Chi, Xiaojiong Chen, Davide Bacco, Qihuang Gong, Leif K. Oxenløwe,  Jianwei Wang, and Yunhong Ding
}

\maketitle

\begin{abstract}
Quantum technology is poised to enable a step change in human capability for computing, communications and sensing.
Photons are indispensable as carriers of quantum information---they travel at the fastest possible speed and readily protected from decoherence.
\ch{However, the system requires thousands of near-transparent components with ultra-low-latency control.
For quantum technology to be implemented, a new paradigm photonic system is required: one with in-built coherence, stability, the ability to define arbitrary circuits, and a path to manufacturability.}
Silicon photonics has unparalleled density and component performance, which, with CMOS compatible fabrication, \ch{place it in a strong position} for a scalable quantum photonics platform.
This paper is a progress report on silicon quantum photonics, focused on developments in the past five years.
We provide an introduction on silicon quantum photonic component and the challenges in the field, summarise the current state-of-the-art and identify outstanding technical challenges, as well as promising avenues of future research.
We also resolve a conflict in the definition of Hong-Ou-Mandel interference visibility in integrated quantum photonic experiments, needed for fair comparison of photon quality across different platforms.
Our aim is the development of scalability on the platform, \ch{to which end we point the way to ever-closer integration, toward silicon quantum photonic systems-on-a-chip}.
\end{abstract}


\IEEEpeerreviewmaketitle

\section{Introduction}
\label{sec:introduction}
\IEEEPARstart{I}n the last five there has been a dramatic acceleration of progress in quantum technology. Propelled by billions of dollars of investment, we have seen loophole free Bell tests~\cite{bell1964einstein, giustina2015significant, shalm2015strong, hensen2015loophole}, quantum communications over thousands of kilometers via satellite link~\cite{ren2017ground, liao2017satellite,  liao2018satellite}, and a quantum information processor outperform the world's most powerful supercomputer by orders of magnitude (at one specific task)~\cite{arute2019quantum}.
Furthermore, quantum enhanced precision is now enhancing the sensitivity in the detection of gravitational waves~\cite{abbott2016observation, aasi2013enhanced, tse2019quantum}---one of the last decade's most significant scientific results.
With the exception of ref.~\cite{arute2019quantum}, photonics is at the core of these technologies.

\ch{
Scaling up quantum systems, that is, increasing computational space, communication distance and measurement sensitivity, is key.
In the last decade, integrated quantum photonics has developed as one potential route to scalability~\cite{politi2009integrated}.
Using photonic integrated circuits, composed of lithographically fabricated miniature waveguides, quantum photonic experiments can be implemented in devices on the cm$^2$ scale. 
This culminates in a reduction size, weight and power by several orders of magnitude  when compared to their counterparts in bulk optics, enabled by miniature and inherently phase stable components~\cite{silverstone2016silicon, wang2019integrated}.
However, many challenges remain, for example limited photon source efficiency and purity, and component losses.
Today, though integrated component losses are steadily improving, the largest quantum photonics demonstrations remain in bulk optics, for example demonstrations of up to 20 simultaneous photons in 60$\times$60 interferometer~\cite{wang2019boson}, 12-photon entanglement~\cite{zhong201812} and 18 entangled qubits~\cite{wang201818}.
}

\ch{While quantum capability has been demonstrated across integrated photonics, silicon quantum photonics is a leader in the scale and scope of its applications.}
Today, silicon photonics provides a versatile testbed for quantum photonic technology, with demonstrations of resources for measurement-based quantum computing~\cite{adcock2019programmable}, high-dimensional entanglement entanglement~\cite{wang2018multidimensional}, robust quantum communications~\cite{cai2017silicon} and photonic quantum machine learning~\cite{carolan2020variational, wang2017experimental} with up to eight simultaneous photons~\cite{paesani2018generation}.
\ch{Furthermore, programmable circuits give access to large classes of quantum photonic capabilities via a single point of optical alignment~\cite{carolan2015universal, shadbolt2012generating, adcock2019programmable}, while complementary metal-oxide-semiconductor (CMOS) compatible fabrication makes electronic co-integration possible~\cite{atabaki2018integrating}.}

Meanwhile, classical silicon photonics leads the way in large-scale photonic integrated circuits.
Classical silicon photonic circuits boast GHz speeds~\cite{chen201522}, up to tens of thousands of optical components~\cite{seok2019240, perez2017multipurpose}, and  electronics and photonics integrated in the same monolithic devices~\cite{thomson2016roadmap,  chung2018monolithically, sun2015single}.
%
%
This places silicon quantum photonics in the position to benefit from the $3\times10^8$ USD a year~\cite{thomson2016roadmap, chen2018emergence} invested in tackling the largest challenges to scaling the classical platform.

At the end of last year a quantum device outperformed all traditional computers~\cite{arute2019quantum}, marking 2020 as the beginning of the era of so-called noisy intermediate scale quantum (NISQ) technology~\cite{preskill2018quantum}.
\ch{Photonic quantum information processors are closely following, with over 20 input photons in 60 modes demonstrated~\cite{wang2019boson}, well into the NISQ domain.}
\ch{In this era, the development of rewarding medium-scale applications is foremost, and will light the path to fault-tolerance (i.e.~a system able to continuously operate despite its errors).}
Many of these applications, such as quantum simulators~\cite{aspuru2012photonic, mcardle2020quantum, sparrow2018simulating}, machine learning~\cite{steinbrecher2018quantum} and graph-based computation~\cite{bromley2020applications, yamamoto2017coherent} have a natural \ch{implementation by quantum photonics.}
While the race to build a scalable fault-tolerant quantum computers has many contenders, the silicon photonic route shows promise~\cite{rudolph2017optimistic, gimeno2015three}: today there are two well-funded companies developing silicon-based photonic quantum computers~\cite{psiq, xanq}.
The ultimate challenges are formidable: to enter the regime of fault tolerance, photon indistinguishability, transmission, and quantum process fidelities must be pushed beyond $0.99$~\cite{rudolph2017optimistic, morley2018loss}.

Meanwhile, photonics is seen as the only viable choice for quantum communications, which promise security beyond what is possible classically~\cite{pirandola2019advances}.
\ch{So far, realisations of satellite-based quantum communications utilised bulk optical transmitters and receivers, developed to extremely robust, space qualified standards~\cite{ren2017ground, liao2017satellite, liao2018satellite}. 
Attaining the same levels of reliability with a system based on integrated photonics remains a challenge, however the potential benefits in size, weight and power---which are particularly costly in space---are large.}
Silicon photonics provides a cost-effective and scalable solution for terrestial quantum key distribution (QKD) nodes, as well as potentially enabling large-scale quantum network tasks in the future, for example by the preparation and distribution of large entangled states~\cite{azuma2015all, markham2008graph}.

In this paper, we give an update on the state of the art of silicon quantum photonics.
As a foremost reference, we point to ref.~\cite{silverstone2016silicon}, which provides an excellent introduction to the field up to 2016.
More general reviews of integrated quantum photonic platforms can be found in refs.~\cite{wang2019integrated, flamini2018photonic}, while hybrid integration and solid-state emitters are adressed in refs.~\cite{elshaari2020hybrid, kim2020hybrid, senellart2017high}.

The review is structured as follows: Sec.~\ref{sec:components} details the state-of-the-art in silicon quantum photonic components---passives, modulators and detectors.
Sec.~\ref{sec:source} introduces photon pair generation in silicon waveguides and reviews recent developments in photon pair source engineering for spectral purity and heralding efficiency\ch{, as well as recent results in the integration of solid-state single photon emitters with silicon-based waveguide circuits.}
In Sec.~\ref{sec:processing} we discuss processing quantum information in silicon photonic circuits and review large-scale and multiphoton capability as well as applications in machine learning and boson sampling, towards a linear optical quantum computer.
Sec.~\ref{sec:comms} discusses silicon photonics for quantum communications technology, focusing on the achievements of a wide variety of communications protocols, and an outlook for future quantum networks.
In Sec.~\ref{sec:sensing} we highlight silicon as an enabling technology for quantum sensing applications.
Finally, in Sec.~\ref{sec:scaling} we give an outlook the future of the platform and provide a list of research programs for today's challenges which will enable future scaling, before concluding.

\section{Components}

\label{sec:components}

\begin{table}[t!]
\begin{adjustbox}{center}

\begin{tabular}{P{4cm}	P{1.5cm}   P{0.7cm}  P{0.7cm} }
Component	& 	Loss (dB)  & (\%)	  &    Ref.			\\
\noalign{\vskip 1mm}    
\hline
\hline
\noalign{\vskip 1mm}
Edge coupler (SMF)	        &	$0.7$      & 	$85$\%  & \cite{cardenas2014high}\\
Edge coupler (tapered fiber)&	$0.3$      & 	$93$\%  & \cite{PU20103678Taper}\\

\noalign{\vskip 1mm}
Grating coupler (1200 nm)    	    &	$0.36$     &  	$92$\%  &\cite{notaros2016ultra}\\
Grating coupler (metal mirror)    	    &	$0.5$     &  	$89$\%  & \cite{HoppeJSTQE20GCmirror}\\
Grating coupler (260 nm si)             &	$0.9$     &  	$81$\%  & \cite{marchetti2017high}\\
Grating coupler (full-etch PhC)         &	$1.7$     &  	$68$\%  & \cite{Ding2013PhCGC}\\
3D vertical coupler          &	$1.0$     &  	$79$\%  & \cite{Luo20OLcoupler}\\

\noalign{\vskip 1mm}
Waveguide, Si (m$^{-1}$)                  &	$2.7$     & 	$54$\%  & \cite{Biberman2012Siwg}\\
Waveguide, TFLN (m$^{-1}$)                &	$2.7$    &  	$54$\%  & \cite{Zhang2017LN}\\
Hybrid Si/SiN delay line (m$^{-1}$)      &	$0.12$   &  	$97$\%  & \cite{Puckett2019SiN}\\
Hybrid Si/silica delay line (m$^{-1}$)   &   $0.037$    &  	$99$\%  & \cite{Lee2012NCSiO2wg}\\

\noalign{\vskip 1mm}
$2\times2$ MMI 3 dB coupler            &	$0.2$    &  	$96$\%  & \cite{Dumais2016OFCmmi}\\

\noalign{\vskip 1mm}
Crosser   &	$0.02$	&   $99.5$\%  & \cite{Zhang2013blochCross}\\

\end{tabular}
\end{adjustbox}
\caption{\textnormal{State-of-the-art losses of silicon photonic components. Thermo-optic and Pockels-based Modulator losses can approach linear waveguide losses by sufficiently distancing metal elements from the waveguide mode. Similarly, losses in $2\times2$ directional couplers follow linear waveguide loss: for $100$ $\upmu$m length, losses of less than $0.003$ could be possible~\cite{Biberman2012Siwg}. Likewise, filter losses can be extrapolated as combinations of these components.}}
\label{tab:losses}
\end{table}

\subsection{Passive components}
\label{sec:passivecomps}

Maximising photon transmission is paramount to the scaling of quantum photonics.
For this reason, scaling quantum information processing in silicon photonics relies on the development of ultra-low-loss passive components.
These include fiber-to-chip couplers, $2\times2$ couplers, cross intersections, multimode and polarisation components, pump rejection filters, and low-loss delay-lines. 
Scaling quantum photonic systems requires the utmost performance from these components, with minimising loss the top priority.
Table~\ref{tab:losses} contains a list of state-of-the-art losses for these components.

\begin{figure*}[ht!]
    \includegraphics[width=1\textwidth]{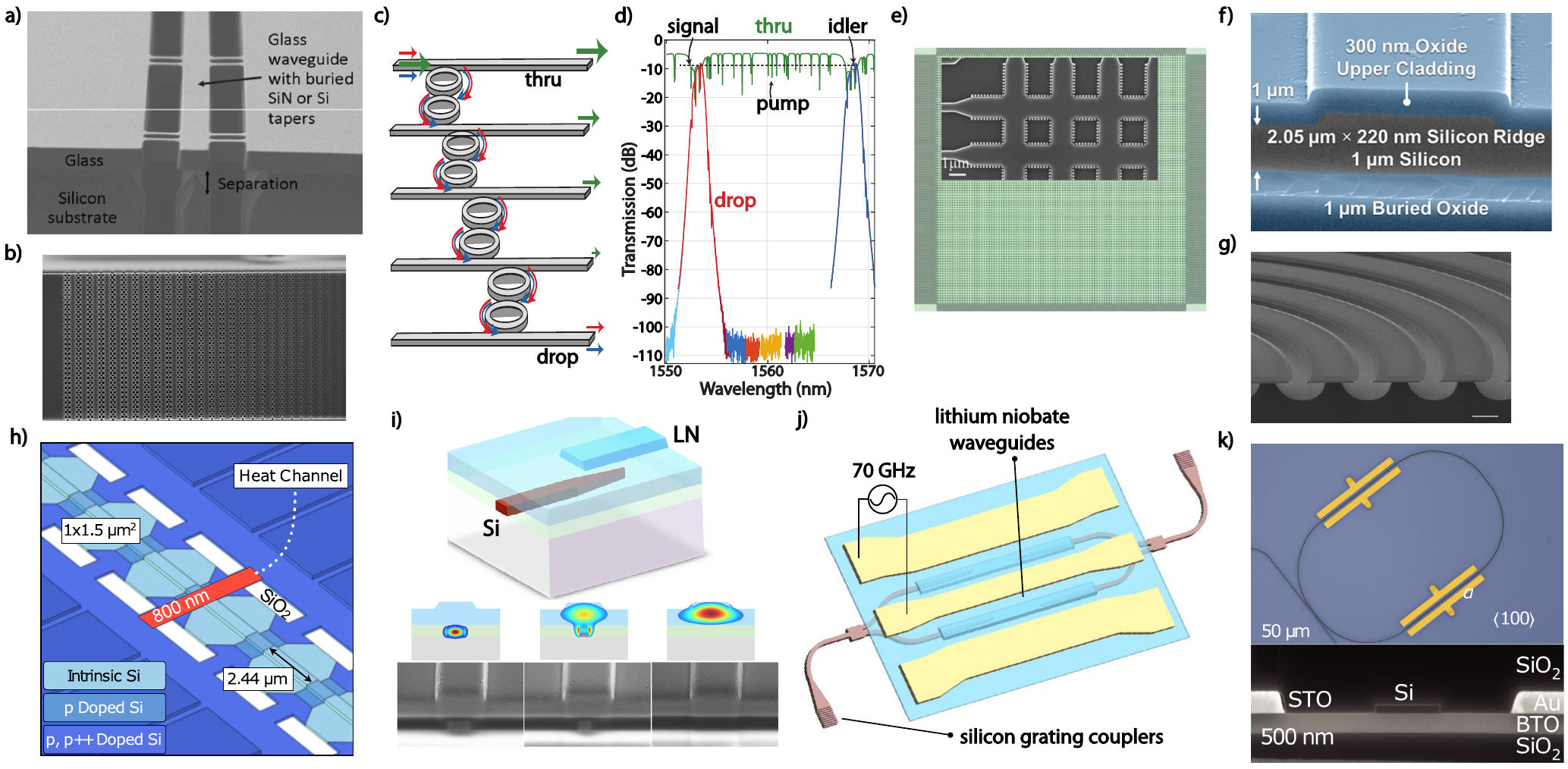}
    \centering
    \caption{Components for silicon quantum photonics. (a) Low-loss cantilever coupler~\cite{Chen2010Coupler} and (b) detail of a photonic crystal grating coupler~\cite{ding2014fully} \ch{(waveguide runs to the right)} for coupling with standard single mode fiber. \ch{The photonic crystal modulates the effective index of the waveguide to create a grating, which directs the light out of plane. A metal mirror on the underside of the grating coupler ensures high-transmission to single mode fiber.}  (c) Four second-order ring resonator filters \ch{demultiplex the single photons (blue, red) from the bright pump light (green) used to generate them via spontaneous four wave mixing}. (d) \ch{Transmission spectrum of the filters of (c) showing 95 dB of pump rejection, so that single photons may be detected~\cite{gentry2018monolithic}}, \textcopyright~The Authors. (e) An ultra-low-loss cross intersection fabric~\cite{Zhang2013blochCross}. \ch{Here, engineering of subwavelength nanostructures which support Bloch waves are used to reduce the insertions loss to 0.01 dB per crossing}. (f-g) Extremely low-loss waveguides based on (f) \ch{shallow-etched silicon ridge waveguide~\cite{Biberman2012Siwg} ac hiving a loss of 2.7 dB/m by minimising side-wall scattering and (g) silica edge mode waveguide~\cite{Lee2012NCSiO2wg} achieving 0.04 dB/m loss. Here, the optical mode is contained within the edge of the silica wedge, which is air clad and supported by a silicon pillar. Etch duration is optimised for smooth silica surfaces, minimising side-wall scattering.} (h) Detailed structure of an optimised through-waveguide thermo-optic phase modulator with 130 KHz bandwidth ~\cite{harris2014efficient}. (i) \ch{Low-loss mode conversion between a silicon waveguide and a thin-film lithium niobate waveguide is achieved via an inverse taper. The lower figure shows three modal cross sections at the beginning, middle and end of the taper}. (j) This structure formed part of a 70~GHz hybrid modulator~\cite{He19NPLNmod}.  (k) \ch{Hybrid silicon-BTO integration~\cite{Abel19BTO}. Here, a silicon waveguide is fabricated on top of a layer of BTO to form a ring resonator. The large pockels coefficient of BTO enables modulation the the phase in the ring with a bandwidth of 30~GHz. The lower inset shows the cross-section of the BTO and Si hybrid integration.}}
    \label{fig:passive}
\end{figure*}

Fiber-to-chip couplers are essential, as their loss significantly restricts the scale of today's on-chip quantum photonic systems.
Silicon devices primarily use one of two kinds of coupler: lateral couplers based on spot-size converters (SSCs, Fig.~\ref{fig:passive}a)~\cite{Chen2010Coupler} , and vertical couplers based on sub-wavelength gratings~\cite{Taillaerts2004GC}. 
SSCs have the advantage of low polarisation dependency and large coupling wavelength bandwidth and are usually based on single~\cite{Chen2010Coupler,PU20103678Taper} or dual inverse tapers~\cite{Wang2016doubletip}. 
Realizing low-loss SSCs requires detailed structures based on complicated fabrication process, as well as excellent die cleaving. 
Less than 1 dB coupling loss has been reported for coupling with lensed fibers~\cite{PU20103678Taper,Ben10PTLcoupler} and single mode fibers~\cite{Chen2010Coupler}.
Further decreasing the coupling loss is challenging since the length of the SSCs---typically a few hundred micrometers long---where sidewall roughness from imperfect fabrication causes scattering loss.

Grating couplers are widely used in integrated photonics as their vertical coupling nature supports wafer scale testing. 
Grating couplers can be designed with long adiabatic taper~\cite{Ding2013PhCGC}, but can also feature focusing for a compact footprint~\cite{Van2007FocusGC}. 
For both types, light scattering from sidewall roughness is nontrivial. 
Standard grating couplers are typically polarisation dependent and bandwidth limited, restricting the use of these degrees of freedom to encode quantum information.
One potential solution is two dimensional grating couplers, which are able to couple orthogonal polarisation of light into separate waveguides (Fig.~\ref{fig:multiph}b)~\cite{wang2016chip,llewellyn2020chip}, however, these structures suffer from increased loss. 
Efficient grating couplers have historically been a challenge due to power leakage in the substrate~\cite{Ding2013PhCGC, Vermeulen2010lowlossGC, ChenPTL2010lowlossGC}.
Further reducing the loss requires directed emission from the grating, which has been demonstrated with distributed Bragg reflectors~\cite{Nambiar2019GCDBR} and metal layers underneath the grating (Fig.~\ref{fig:passive}b)~\cite{ding2014fully, Zaoui2014GCmirror, Benedikovic2015GCmirror, HoppeJSTQE20GCmirror}, achieving down to $0.5$ dB loss.
However, the addition of a metal mirror is a specialised processing technique with limited availability and process compatibility. 
Other solutions for directing waveguide emission are being developed, for example $0.9$ dB loss has been achieved by optimising the apodisation of fill factor and grating period, as well as the waveguide height and partial etch depth~\cite{marchetti2017high}.
Ref.~\cite{notaros2016ultra} exhibits an impressive $0.36$ loss dB grating coupler on a commercial 45 nm process with HSQ-topped waveguides and two partial etch depths, using a Bloch-Floquet band-structure type optimisation
\cite{notaros2015band}.
Finally, a 3D polymer coupler showing coupling loss of 1~dB was recently demonstrated~\cite{Luo20OLcoupler}.

Mach-Zehnder interferometers (MZIs) are a fundamental to manipulating light in silicon photonics, and consist of two modes joined by two $2\times2$ couplers (either multimode interferometers or directional couplers) with an enclosed phase modulator, forming a programmable interferometer. 
Identifying the two modes with the $\{\ket{0},\ket{1}\}$ basis states of a qubit, MZIs equipped with three phase modulators implement arbitrary single qubit gates (Fig.~\ref{fig:plex}b).
MZIs with mismatched internal path-lengths result in a sinusoidal wavelength filter, useful for splitting single photon wavelengths.
As a choice of $2\times2$ coupler, MMIs have the advantage of large fabrication tolerance, and low insertion losses of less than $0.2$ dB has been reported~\cite{Dumais2016OFCmmi}.
Directional couplers---typically tens of microns long---can exhibit reduced loss when compared to MMIs, typically only suffering from standard waveguide propagation losses, i.e.~from sidewall roughness.
However, they also exhibite increased fabrication sensitivity as they rely on small, 200-500 $\upmu$m gap between the waveguides, to which the coupling and bandwidth is exponentially sensitive to, though it has been shown that this sensitivity can be alleviated by subwavelength structures~\cite{Halir2012ColorlessDC, Xie2019GratingDC}.

In silicon waveguides, polarisation and spatial mode are both available carriers of quantum information, with coherent conversion---recently demonstrated~\cite{feng2016chip}---enabling new functionality, for example interfacing with fiber networks and dense information coding.
Polarisation multiplexers can be realised by polarisation dependent directional couplers which exploit the differing coupling constants of TE and TM modes, leading to bandwidths which can be more than 100~nm~\cite{Fukuda06PBS}.
Mode (de)multiplexers can be also realised by directional couplers~\cite{Dai2012SiModeMUX,Wang2014MDM8}. 
So far, up to eight modes, including two polarisations, have been demonstrated~\cite{Wang2014MDM8}.
Here, the cross-mode coupling is sensitive to fabrication tolerance, though a taper in the coupling region alleviates this somewhat~\cite{Ding2013TaperDC}.

In quantum photonic devices, pump light used to generate single photons must be removed before those single photons can be detected. 
Today, this is typically achieved with off-chip components.
However, future integration of single-photon detectors necessitates on-chip pump removal.
Over 100 dB of extinction---with minimal losses in the single photon bands---is required.
Similar filters can also be used to separate classical and quantum channels in quantum communications networks.
To this end, cascaded microring resonators forming coupled resonator optical waveguides (CROWs)~\cite{Ong2013PTLfilter} and cascaded asymmetric MZIs~\cite{Piekarek2017filter} have been used to achieve more than 50~dB extinction on a single chip. 
Here the filtering ratio was limited by pump light scattered into the chip substrate---particularly significant due to the low $-5$ dB efficiency of the grating couplers used. 

In 2018 single photon generation and pump rejection were demonstrated together on a single chip for the first time~\cite{gentry2018monolithic}. Here, four second-order ring resonator filters isolates the signal and idler bands from pump (Fig.~\ref{fig:passive}c,d).
Other recent results include cascaded grating-assisted contra-directional couplers with extinction ratio more than 70~dB in silicon nitride~\cite{Nie2019filter}, and a third-order ring resonator with extinction ratio more than 80~dB~\cite{Huffman17arxivfilter} (also in silicon nitride), and the use of two chips to produce more than 100~dB of extinction~\cite{Ong2013PTLfilter, Piekarek2017filter, Kumar2020OLfilter}.
Meanwhile, increases to source efficiency lessen the requirement~\cite{ma2020progress}.

In quantum photonic circuits, it is often unavoidable to have waveguides cross one another, particularly when using path-encoded qubits~\cite{wang2017high}.
Here, crosstalk can produce qubit errors, and thus is a key metric.
Promising waveguide crossers based on MMI structures~\cite{wang2018multidimensional, Chen2006CrossMMI}, shaped waveguides~\cite{Fukazawa2004cross, Sanchis2009cross, Ma2013cross}, tilted waveguides~\cite{Xie2011SPIE,Kim2014tilt}, subwavelength gratings~\cite{Bock2010sub}, and by inverse-design~\cite{Yu2019InverCross}, have been reported, with typical insertion losses of less than $0.2$ dB.
Today, Bloch mode structures exhibit the best performance, with sub $0.03$ dB loss (Fig.~\ref{fig:passive}e)~\cite{Zhang2013blochCross, Liu2014blochCross}, and $-35$ dB crosstalk or less. 
Refs.~\cite{wu2020state} and \cite{chang2020silicon}, published in 2020, review waveguide crossers.

Typical propagation losses of silicon nanowaveguides are around 2~dB/cm, which is prohibitive for scaling quantum applications. 
Thanks to developments in fabrication technology such as oxidization~\cite{Lee2001lowloss,Sparacin2005oxi}, etchless waveguide fabrication~\cite{Cardenas2009etchless}, Hydrogen thermal annealing~\cite{Bellegarde2018HeAnneal}, and shallow waveguides~\cite{Dong2010shallow,Biberman2012Siwg}, propagation loss has been significantly reduced, with a record-low loss of $2.7$~dB/m (Fig.~\ref{fig:passive}f)~\cite{Biberman2012Siwg}.
Reducing propagation loss further will require these sidewall smoothing techniques to be combined~\cite{Lee2001lowloss}. 
Hybrid integration with silicon nitride waveguide or small-angle wedge silica waveguides are other options under development, with extremely low propagation loss of 0.123~dB/m~\cite{Puckett2019SiN} and 0.037~dB/m~\cite{Lee2012NCSiO2wg} demonstrated, respectively (Fig.~\ref{fig:passive}g). 
Ultra-low-loss delay lines are a critical component for (de)multiplexing technology~\cite{kaneda2019high, Lenzini2017deMUX, Hummel2019deMUX}.
Delay lines are discussed further in Sec.~\ref{sec:determinism} in the context of multiplexing for deterministic single photon sources.

\subsection{Optical modulators and switches}
\label{sec: modulators}

Today, thermo-optic phase modulators (Fig.~\ref{fig:passive}h) \cite{fang2011ultralow, harris2014efficient}  remain the only low-loss, small-footprint, phase shifters available silicon photonics.
These can be implemented with $p$ or $n$ doped silicon connecting with the waveguide~\cite{watts2013adiabatic} or with metal heating elements suspended close to the waveguide~\cite{fang2011ultralow}. 
Though dopant-based thermo-optic modulators can have larger bandwidths due to their proximity with the waveguide, the charge carriers introduced induce excess loss.
Furthermore, thermo-optic modulators have maximum operating speed in the kilohertz range, placing severe restrictions on their application to feed-forward and multiplexing techniques, needed for scalable quantum photonics.
Modulators and switches based on plasma dispersion effect are attractive due to the CMOS compatible fabrication process and high-speed up to tens of GHz~\cite{Reed2010SiMod}. 
However the plasma dispersion effect effect is weak and results in low efficiencies, large device footprint and power consumption.

Modulators based on the Pockels effect are more promising, thanks to their low loss and femtosecond response time~\cite{Li19NMPockels}. 
Thin-film lithium niobate (LiNbO$_3$, TFLN) and titanate (BaTiO$_3$, BTO) modulators and switches have gained significant attention.
Historically, fabricating TFLN waveguides is a challenge, but thanks to recent progress in manufacturing~\cite{levy1998fabrication, hu2007lithium} it is now emerging as a desirable integrated photonics platform~\cite{rabiei2013heterogeneous, poberaj2012lithium, bazzan2015optical}, with state-of-the-art loss at $2.7$ dB/m ~\cite{Desiatov2019LN, Zhang2017LN}. 
Meanwhile, high-speed, low-loss and low-driving voltage modulators and switches have been demonstrated~\cite{Wang2018LNmod, Zhang2018OFCLN, wang2018integrated, RaoJSTQE2018LN}, and low-loss integration to silicon photonics (via wafer bonding) has been shown with negligible decrease in performance~\cite{He19NPLNmod, Sun2020OFCLN, Gao19PTLLNswitch}, making this platform very attractive for quantum applications (Fig.~\ref{fig:passive}i-j).
For example, phase modulation rates of over $70$ GHz have been demonstrated with hybrid TFLN ridge waveguides, where the silicon to TFLN transition is achieved with a low-loss taper \cite{He19NPLNmod, Sun2020OFCLN}, resulting in a switching voltage length product of $V_\pi L = 2.2$ Vcm.
In TFLN, high-efficiency fiber coupling is a challenge, whereas hybridisation with silicon---where efficient light coupling options exist---can be achieved with virtually zero loss using inverse tapers (Fig.~\ref{fig:passive}i).
Similar results have been achieved with silicon nitride waveguides in ref.~\cite{boynton2020heterogeneously}, where the authors hint at heterogeneous integration with CMOS electronics.
Indeed, TFLN on silicon presents an attractive route to fast-clock-rate electronic-photonic co-integration, with CMOS-voltage compatible switching already demonstrated~\cite{wang2018integrated} alongside a $2$ cm MZI switch with $0.5$ dB loss.
BTO, with its ultra-high Pockels coefficient, is another attractive option, and BTO-Si hybrid modulators have been demonstrated up to 50 GHz (Fig.~\ref{fig:passive}k)~\cite{Abel19BTO, Eltes2019JLTBTO}, and at cryogenic temperatures~\cite{eltes2019integrated}.
However, BTO is typically amorphous, resulting in larger excess loss, and currently has poor fabrication availability. 
In silicon nitrite, similar results have been obtained using lead zirconate titanate (PZT)~\cite{Alexander18SiNPockels}.

Finally, it is possible to equip pure silicon with a $\chi^{(2)}$ by inducing symmetry breaking deformations to its lattice structure~\cite{jacobsen2006strained}.
Recently the DC Kerr effect was used to modulate phase at up to $1$ GHz in silicon waveguides~\cite{Timurdogan17SiPockels}, also at cryogenic tempertures~\cite{chakraborty2020cryogenic}.
Here, $p$-$i$-$n$ junctions create strong electric fields to distort the crystal lattice, breaking its symmetry---though free carrier absorption resulted in a phase-dependent loss of up to $1$ dB/cm.

\begin{figure}[t!]
    \includegraphics[width=.48\textwidth]{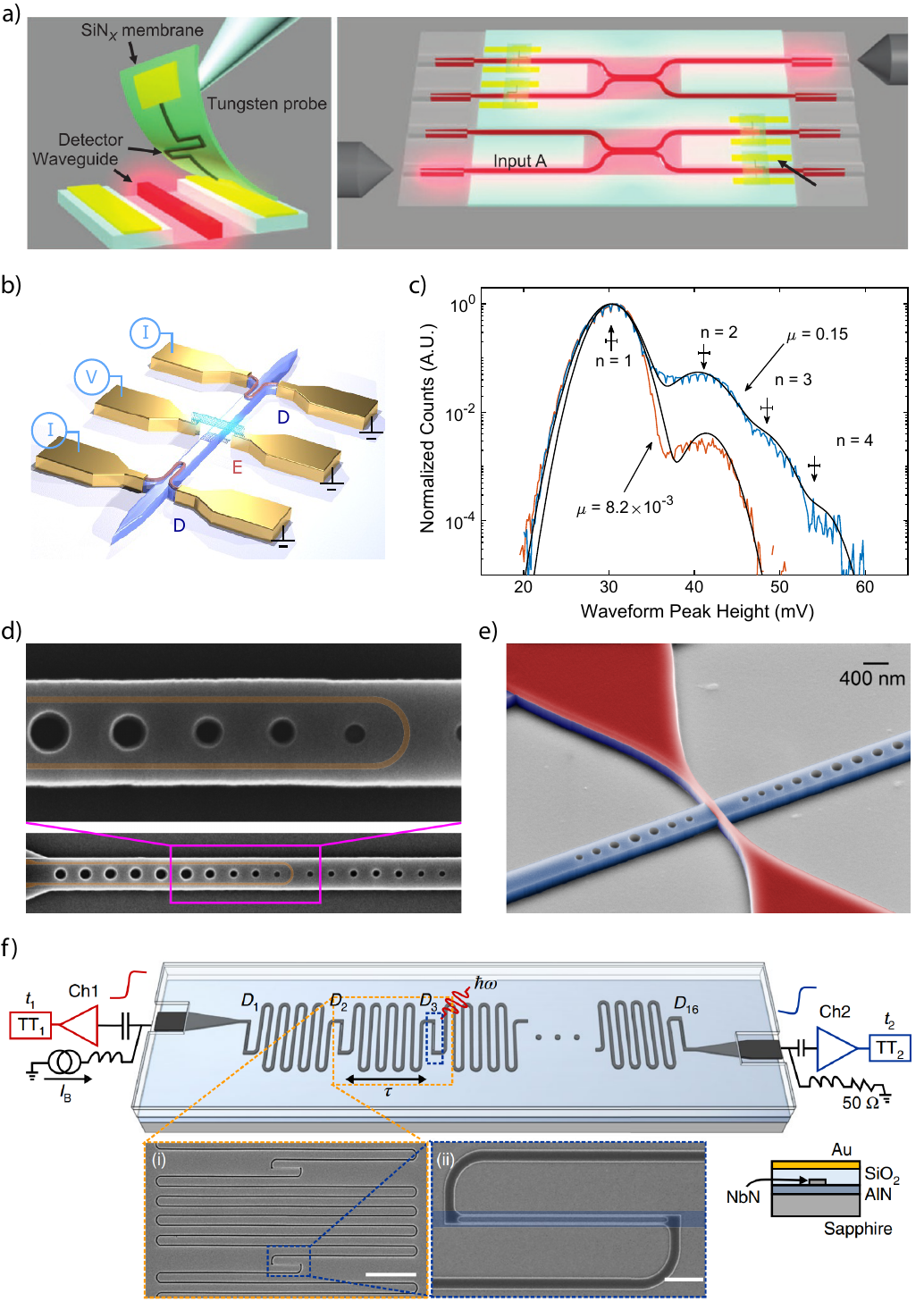}
    \centering
    \caption{Integration and scalability of SNSPDs. (a) A silicon nitride wafer with an SNSPD is bonded to a rudimentary silicon waveguide circuit \ch{by micromanipulation}~\cite{najafi2015chip}. (b) An integrated Hanbury-Brown and Twiss experiment on silicon using an electrically driven carbon nanotube photon source and waveguide integrated SNSPDs~\cite{khasminskaya2016fully}. (c) Photon number resultion in a standard SNSPD by analysing peak height; $\mu$ is mean photon number~\cite{cahall2017multi}. (d,e) SNSPDs integrated in a photonic crystal cavities decrease recovery time, increase wavelength selectivity, and decrease form factor~\cite{akhlaghi2015waveguide, vetter2016cavity}. (f) Signal multiplexing and number resolution via pulse timing analysis in an row of of SNSPDs wired in series~\cite{zhu2018scalable}.}
    \label{fig:detectors}
\end{figure}

\subsection{Single photon detectors}
\label{sec:dets}

Single photon detectors are an essential component of photonic quantum technologies. 
Opaque below around $1100$ nm, silicon photonics are typically optimised for telecommunication bands, typically at around $1530$--$1565$ nm to align with the maxima in single mode fiber transmission.
At these wavelengths, there are limited options for single-photon detectors, though InGaAs detectors offer a relatively inexpensive solution~\cite{Namekata09APD, Zhang15SPD}.
However, the efficiency of InGaAs detectors is typically lower than 10\%. Maximum count rates can be measured in MHz and the dark-count rate is typically a few hundred Hz.
These find applications in short-medium link quantum communication scenarios where their loss may be budgetable.

Though room-temperature operation is preferable for practicality and cost, superconducting nanowire single photon detectors (SNSPDs), which work at cryogenic temperatures (typically lower than $4$ K) provide excellent performance~\cite{Natarajan2012SNSPD, Rosenberg13SNSPD}.
Systems with efficiencies of up to 95.5\%~\cite{photonspot} are commercially available, exhibiting darkcount rates of less than less than 100 Hz and recovery times which enable more than $10^7$ detections per second~\cite{Zhang17SNSPD}. 
Today's commercial SNSPD systems have enabled much progress in modern silicon quantum photonics experiments.
However, these are necessarily off-chip and fiber coupled---an inherent source of loss.
To move forward, on-chip integration of SNSPD systems is required.

To this end, SNSPDs have been demonstrated on silicon waveguides in a travelling wave configuration~\cite{pernice2012high, schuck2013waveguide} with close to unity efficiency \cite{akhlaghi2015waveguide}, showing great promise for integration.
One impressive demonstration comes from \ch{transferring a membrane of silicon nitride containing} SNSPDs onto silicon waveguides \ch{via micro-manipulation}, which the authors use to perform an on-chip $g^{(2)}(t)$ experiment with an integrated beamsplitter, establishing the photons single-ness (Fig.~\ref{fig:detectors}a)~\cite{najafi2015chip}.
Though less intrinsically scalable than lithographically patterned SNSPDs, this technique enables the addition of SNSPDs to any air-clad silicon circuit.
More recently, detectors have been written on a commercial silicon photonic device, using custom thin-film techniques (Fig.~\ref{fig:QC}c)~\cite{Zhang19arxivChip}, pointing the way to scalable patterning of SNSPDs on silicon waveguides, though further work is needed to improve the $-6$ dB grating coupler efficiency shown.
Silicon waveguide SNSPDs have been also integrated with on-chip light sources: an LED~\cite{buckley2017all}, and a carbon nanotube emitting single photons ($g^{(2)}(0) = 0.49$, Fig.~\ref{fig:detectors}b)~\cite{khasminskaya2016fully}.
These impressive demonstrations are able to show single photon operation by virtue of featuring electrically driven light source, circumventing the need to remove pump light such that the device's detectors are not overwhelmed.

Detector recovery times, usually around 50 ns in commercial systems is a key metric limiting data rates (see Sec.~\ref{sec:determinism}).
Reducing the long detector length---usually required for efficient photon absorption---is one route to faster operation.
Silicon photonic crystal cavities have been used to enhance absorption of a short detector, demonstrating recovery times around 500 ps~\cite{vetter2016cavity, munzberg2018superconducting, akhlaghi2015waveguide} (see Figures \ref{fig:detectors}d,e).
Cavities and can also provide ultra-narrow wavelength specificity which, though not desirable in some applications, may find application in frequency multiplexing.
Similarly, detector jitter (temporal resolution) forms a current limitation in quantum communications, and will form a bottleneck in ultra-fast temporal multiplexing using TFLN or BTO modulators.
Recently, record-low $3$ ps resolution was demonstrated with a $5$ $\upmu$m detector and cryogenic electronics, exposing intrinsic jitter processes of SNSPDs on a silicon substrate~\cite{korzh2020demonstration} .
The low jitter of these SNSPDs, made from RF-bias sputtered niobium nitride films, is found to be due to their quasi-amorphous small-crystal structure, resulting in $2$--$3.5$ times the responsivity of competitor materials such as WSi.

In some applications, such as Gaussian boson sampling~\cite{paesani2018generation} 
and schemes for universal photonic quantum gates~\cite{Bartlett02PRA}, photon number resolving detectors are needed. 
For example, number resolution can be achieved by space- or time-multiplexing SNSPDs~\cite{divochiy2008superconducting, allman2015near, sahin2013waveguide, Fitch03Photonnumber, mattioli2016photon} in a monolithic device.
Though not demonstrated on the silicon photonic platform yet, these techniques rely on processing of the electronic detector pulses, and so are in principle transferable to any platform.
Spectral multiplexing of SNSPDs has also been demonstrated~\cite{kahl2017spectrally}, for example with a broadband chip-scale single-photon spectrometer covering visible and infrared bands~\cite{Cheng2019Broadband}, with applications to frequency-domain source multiplexing.
This device comprises a pulse time-of-flight measurement on a long SNSPD and features a $7$ nm detection resolution and over $200$ wavelength detection channels.
The technique was originally developed in ref.~\cite{zhao2017single} to produce a single photon camera with area 590 effective pixels in $286 \times 193$ $\upmu$m$^2$.
One promising application comes from photon number resolution achieved with a single conventional SNSPD by processing of its output pulses (Fig.~\ref{fig:detectors}c)~\cite{cahall2017multi}.
Though only resolution up to $n = 4$ is shown, this is a simple path to empowering multiplexed photon pair sources---in which $n>3$ events are very unlikely---to discard multiphoton events (see Sec.~\ref{sec:determinism}).
Titanium-based superconducting transition-edge detectors (TESs)~\cite{Fukuda2011photonnumber, calkins2013high} offer the premier solution number-resolution, with excellent discrimination up to 15 photons in real time by pulse-area analysis~\cite{morais2019revealing}.
However, TESs suffer from KHz max repetition rates, limiting their applicability.

\begin{figure*}[t!]
    \includegraphics[width=1\textwidth]{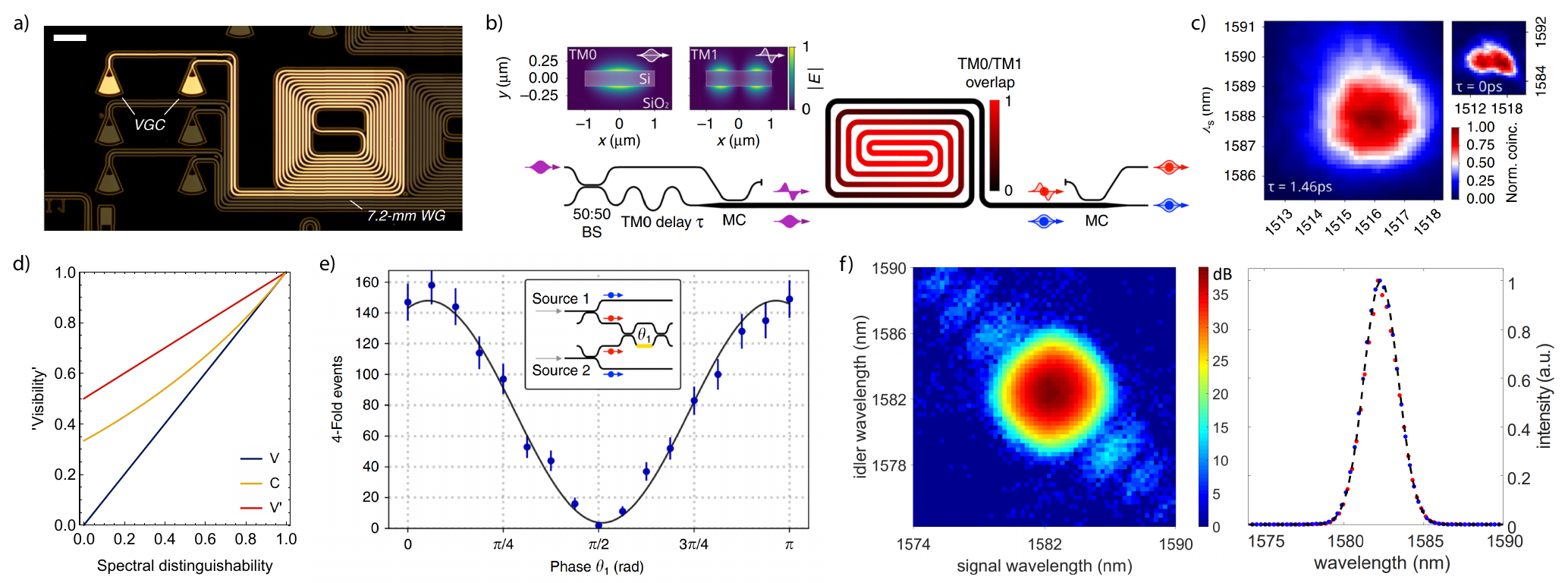}
    \centering
    \caption{Photon pair sources, quantum interference and purity. (a) A simple silicon waveguide source. Here, a 340 nm silicon wafer enables phasematching for photon pair generation at $2.1$ $\upmu$m~\cite{rosenfeld2019mid}, \textcopyright~The Authors. (b) A  multimode waveguide source enables a gaussian phasemathcing profile via the convolution of TM0 and TM1 modes with different group velocities~\cite{paesani2020near}. (c) The JSA of produced by (b) with and without (inset) a group delay between pulses. With delay, the spectral purity was $0.99$ without narrow filtering. (d) Scaling of conflicting definitions of HOM fringe visibility found in the literature. $V$ corresponds exactly to distinguishability and standard HOM dip visibility, while $C$ and $V'$ are overly favourable. (e) Record on-chip heralded HOM interference with $0.96$ visibility in silicon~\cite{paesani2020near}. (f) Joint spectrum and marginal spectra of and of a high spectral purity bulk optical source with near-gaussian shaped nonlinearity, used to produce heralded HOM interference with $0.984$ visibility~\cite{chen2019indistinguishable}. Note the logarithmic scale of the JSA, required to view the suppressed sinc-shaped side lobes.}
    \label{fig:sources}
\end{figure*}

As systems are scaled up, multiplexing the readout of many SNSPDs will be required, due to constrained thermal budget~\cite{gaggero2019amplitude}.
One recent demonstration shows single-channel readout of all $136$ one- and two-photon events across 16 detectors, as well as photon number resolution (Fig.~\ref{fig:detectors}f)~\cite{zhu2018scalable}.
Cryogenic signal processing and multiplexing techniques are undergoing rapid development in the context of SNSPD arrays, where up to $32\times 32$ pixels demonstrated with row-column signal multiplexing~\cite{wollman2019kilopixel}.
\ch{Another promising avenue is to use optical data transfer via fiber out of the cryostat.
Recently, ref.~\cite{de2020photonic} demonstrated a modulator driven directly by a nearby integrated SNSPD---a powerful capability for signal multiplexing as well as high-speed multiplexing technologies.}

One potential alternative to bulky and expensive cryogenic comes from silicon-germanium single photon avalanche detectors, which can be waveguide coupled~\cite{martinez2017single} and have recently been demonstrated to be 38\% efficient at $1310$ nm at 125 K~\cite{vines2019high}.
These devices currently suffer from large dark counts in the $10^4$ range and timing resolution of hundreds of picoseconds.
With further improvements, non-cryogenic operation will be attractive to many quantum applications.

\section{Waveguide photon pair sources}
\label{sec:source}

The quality of photon pair sources determines both the error rates and scalability in of quantum photonic technology.
The ideal photon source is deterministic, pure and indistinguishable.
That is, when activated it will produce a single photon with unity probability and that photon will not be entangled to any other system or be in a superposition of number basis.
Additionally, subsequent photons from the same source will be identical in every degree of freedom but time (indistinguishability).
Finally, it must be repeatable---photons emitted from independent copies of the source must be indistinguishable.
The quality of interference between two photons determines their error rate when used as qubits, and is determined by the visibility $0 < V < 1$ of a heralded Hong-Ou-Mandel interference experiment~\cite{hong1987measurement} with four detected photons\footnote{For true single photon sources, such as solid state emitters, this is a two-photon experiment.}.
Note that two-photon HOM interference between a single photon pair is not sensitive to photon purity~\cite{silverstone2015qubit}.
Heralded HOM visibilities indicate a gate-level errors in quantum photonic information processing, and so for fault-tolerance visibilities of over $0.999$ will be required for fault-tolerant quantum computation~\cite{rudolph2017optimistic}.
In the near term, increased HOM visibilities effectively allow larger numbers of photons to interfere~\cite{renema2018efficient}, for example in boson sampling~\cite{BosonSamplingReview}.
The efficiency of a photon source, that is, the probability $p$ that a photon is emitted to its application, combined its transmission from application through to detector, $\eta$, and repetition rate $R$ determine its scalability---$n$-photon data rate scales as $R \eta^n p^n$ for $n$ sources.

Photon sources in silicon quantum photonics are based on spontaneous four-wave mixing (SFWM), where two photons from a bright pump field are converted into an energy- and momentum-conserving photon pair by the $\chi^{(3)}$ nonlinearity of crystalline silicon waveguides. The photons of this pair are historically named the signal and idler photons.
\ch{The quantum state generated is known as a squeezed vacuum state:
\begin{equation}
\ket{\psi} = \sqrt{1- |\xi|^2} \sum_{n=0}^{N}  (-\xi)^n \ket{n n}_{si} .
\end{equation}
Which is normalised for $N\rightarrow \infty$ and $\xi = i e^{i \mathrm{arg}(\zeta)} \tanh|\zeta|$ for squeezing parameter $\zeta  \propto I^2$ for pump power $I$.
For small $\xi$, $\ket{\psi}$ is dominated by vacuum, with an $O(\xi)$ two-photon component.
Four-photon and above components, which are $\leq O(\xi^2)$, can be made small by controlling the pump laser power.
This state can be approximated as a photon pair source with probability of emission $p \approx \xi^2$, with multiphoton component $O(p^2)$.
This trade off between brightness and multiphoton noise (photon number impurity) is characteristic of nonlinear sources in general, with typical brightnesses in the $0.05 < p < 0.1$ range, depending on application.
In modern quantum photonics experiments, the $n$-photon data rate $R \eta^n p^n$,  is dominated by $p$.}

In silicon quantum photonics, delivery of identical coherent pump pulses to multiple sources is achieved by coupling phase-stable and single-mode waveguides. 
Today, HOM interference visibility is limited by spectral purity and multiphoton contamination (number-basis impurity), with $0.98$ overlapped spectra possible with standard SOI lithographic process~\cite{silverstone2014chip}.
Remaining spectral deviations are caused by fabrication tolerances and can be asymptotically improved by improved lithography.

\ch{The purity $P\in[0,1]$ of a quantum state is a measure of the degree to which a photon is entangled with any other system, and is evaluated as $\mathrm{tr}\rho^2$ for a state with density matrix $\rho$.
Pure states ($P=1$) are completely unentangled, while mixed states $P<1$ have entanglement with another system.
For example, if one photon is spectrally entangled with another photon, it's wavelength is equally determined by its partner system, and therefore is in a mixed state when considered individually---it can not interfere deterministically with another photon and $P = \ovalap{\psi}{1} < 1$.
Equally, we say the photon is \emph{distinguished} from other photons by the entanglement it possesses.}

\ch{In the single photon emitter literature, \emph{purity} is also used to describe the number-basis deviation of a photonic state, where, for for example, the state $\ket{\Psi} = \alpha{\ket{1}} + \beta\ket{2}$ is said to have \emph{purity} $|\alpha|^2 < 1$, since this is a key figure of merit of the single photon emitter.
Note that in the quantum information sense, $\ket{\Psi}$ has Purirty $P=1$
Since single emitters are engineered to produce precisely one photon, spectral entanglement with another photon is not a key issue---though entanglement to the emitter system will appear as distinguishability in a HOM measurment.
Meanwhile in nonlinear photon pair source literature in purity is generally used to refer to a state's (lack of) entanglement.
Instead, terms such as \emph{multiphoton noise} or \emph{contamination} are used to refer to number basis deviation (photon singleness), which is an inherent property of the squeezed vacuum state from which photon pairs are generated, as discussed above, and is as such well-understood.
}

HOM visibility quantifies spectral distinguishability, number basis deviation (multiphoton terms), as well as unwanted entanglement of the photons, and as such is the gold standard determining photon quality.
However, heralded HOM experiments are time consuming and resource intensive, and do not indicate the cause of the lack of interference.
When developing single photon sources, rapid characterisation of their properties is a must. 
\ch{The number-basis deviation of a photonic state is established by the second order correlation function $g^{(2)}(t)$, typically in a Hanbury-Brown and Twiss measurement.
A $g^{(2)}(0) = 0$ indicates a single photon state, but is not sensitive to impurity from unwanted entanglement.
For nonlinear photon pair sources, a heralded $g^{(2)}(t)$ is measured, while an unheralded $g^{(2)}(t)$ measurement (without filtering) obtains the spectral purity of the source~\cite{christ2011probing} (for non-resonant sources).}
Meanwhile the magnitude (but not phase) of photon pair source's joint spectral amplitude (JSA, Fig.~\ref{fig:sources}c,f), which dictates its spectral purity, can be rapidly obtained by stimulated four-wave mixing~\cite{eckstein2014high, silverstone2015qubit}, or chromatic group dispersion time-of-flight~\cite{avenhaus2009fiber} measurements.
\ch{Recent extensions of this technique have enabled measurement of the phase, gaining knowledge of the full JSA~\cite{jizan2016phase}.
This was achieved by engineering a coherent phase reference beam with which to interfere the output of the stimulated emission tomorography.
Similarly, the JSA of a silicon ring resonator was recently measured using straight waveguide as a nonlinear phase reference~\cite{borghi2020phase}.}
Finally, the spectral overlap of two photon pair source's JSAs corresponds exactly to the contrast of reverse HOM (rHOM) interference~\cite{paesani2020near}.

\subsection{Quantifying quantum interference}

Recently, there have been examples of conflicting definitions of heralded HOM interference visibility in the integrated photonics literature~\cite{faruque2018chip, adcock2019programmable, vergyris2016chip}.
We will now de-mystify this phenomena, and derive an expression which is both physically motivated and corresponds values given in other quantum photonic experiments.

In a standard HOM interference experiment, two photons are incident on a beamsplitter. The time of arrival of one of the photons is adjusted to find the point of synchrony, where photon coincidence and the output of the beamsplitter is minimised due to maximised quantum interference.
In this traditional HOM experiment, the visibility is defined:
\begin{equation}
    V_{\mathrm{HOM}}= \frac{a-b}{a}
\end{equation}
Where $a$ is the coincidence reference level (far from synchrony), and $b$ is the coincidence minimum.
Far from synchrony, the photons scatter probabilistically, and so the reference level corresponds to half the rate of experimental trials.
Importantly, for single photons, $V_{\mathrm{HOM}} = |\ovalap{\psi_1}{\psi_2}|^2$, the overlap of the quantum states of the photons~\cite{adcock2019generating}, which is the level of indistinguishability in all degrees of freedom bar time-of-arrival, which has been fine-tuned
This quantity can be interpreted as the probability---upon measurement---that the photons interfered at the beamsplitter.

However, in integrated photonics equal time-of-arrival is naturally achieved by path-length matched waveguides, and so HOM interference can be measured in an MZI, which acts as a tunable beamsplitter.
Here, in the swap or bar configuration of the MZI, all experimental trials lead to coincidence.
Thus the maxima of the fringe experiment corresponds to twice the reference level of the dip experiment.
Meanwhile, the minima of the fringe occur when the MZI implements a $50$:$50$ beamsplitter, and is equivalent to the minimum of the dip.
Therefore in a HOM experiment using an MZI, the following expression should be used:
\begin{equation}
    V_{\mathrm{HOM}} =\frac{\alpha-2\beta}{\alpha} = \frac{a-b}{a}
\end{equation}
Where $\alpha = 2a$ is the fringe maximum and $\beta = b$ is the fringe minimum (when tuning the MZI phase)~\cite{adcock2019programmable}.

Some examples in the literature use a different expression to quantify HOM interference in a MZI, which we name `contrast' for ease of discussion:
\begin{equation}
    C = \frac{\alpha-\beta}{\alpha+\beta}
\end{equation}
Both definitions of visibility, as well as the naive $V' = \frac{\alpha-\beta}{\alpha}$ are a monotonic figures of merit of quantum interference, however, neither $C$ and $V'$ appear physically motivated.
In fact, for completely distinguishable photons, $V' = \nicefrac{1}{2}$ and $C = \nicefrac{1}{3}$, whereas $V_{\mathrm{HOM}} = |\ovalap{\psi_1}{\psi_2}|^2 = 0$, though $C$ does converges to $V_{\mathrm{HOM}}$ for highly indistinguishable photons.
Fig.~\ref{fig:sources}d shows the scaling of $V$, $C$ and $V'$ with distinguishability
$C$ finds use in other source characterisations: the contrast of a two-photon rHOM fringe corresponds is equal to the overlap of the two source's JSAs~\cite{paesani2020near}.

\subsection{Source geometry and photon purity}

Repeatable sources of SFWM have been paramount in the emergence of large-scale silicon quantum photonic circuits.
A length of waveguide is all that is required to produce photon pairs---as demonstrated in the first demonstration in 2006~\cite{sharping2006generation}.
A staticly-tuned filter, for example a ring resonator or asymmetric MZI, can be used to split the non-degenerate photons into two qudits, while degenerate photons (generated using two pumps in what would otherwise be the signal and idler channels) can be split by rHOM interference~\cite{silverstone2015qubit}.
However, simple waveguide sources have severely limited spectral purity.
Energy conservation (which asserts $\omega_{p_1}+ \omega_{p_2}= \omega_{s}- \omega_{i}$) causes the photons to be strongly anticorrelated in frequency, though a wavelength-broad pump \ch{relaxes} this constraint, introducing width to the band of anticorrelation.
These correlations are spectral entanglement between the photons and is quantified by photon spectral purity $0 < P < 1$ (an single, unentangled, photon is pure).

Momentum conservation (which asserts  $k(\omega)_{p_1}+ k(\omega)_{p_2}= k(\omega)_{s} +k(\omega_{i})$) produces the phasematching component of the the JSA, which can be used to tame the correlations from energy conservation.
For example, a recent innovation is to implement a gaussian shaped nonlinearity via engineering the domains in a periodically-poled potassium titanyl phosphate (ppKTP) crystal.
This technique was recently demonstrated in bulk optics, achieving $V_\mathrm{HOM} = 0.984$ (Fig.~\ref{fig:sources}f)~\cite{graffitti2018independent, chen2019indistinguishable}.
In contrast, the usual, `top-hat' nonlinearity produced by a discrete nonlineariy produces sinc-shaped JSA, and typically acts to reduce spectral purity.

Straight waveguide sources are often used in combination with off+chip pump rejection filters to isolate a symmetric portion of a JSA, leading to an apparent increase in spectral purity, and interference visibility.
Typically, fiber filters (for example designed for wavelength division multiplexing) are used to isolate channels close to the pump, with $V_\mathrm{HOM} = 0.80$ demonstrated via $0.7$ nm filters with $1.8$ nm wide pump pulses~\cite{adcock2019programmable}.
However, this filtering has fundamental disadvantages: firstly the crucial photon rate is reduced, and secondly the presence of one photon can no longer be used to herald (guarantee the presence of) its partner.
This reduction in heralding efficiency implies a strict limit on the effectiveness of source multiplexing.

\begin{figure}[t!]
    \includegraphics[width=0.48\textwidth]{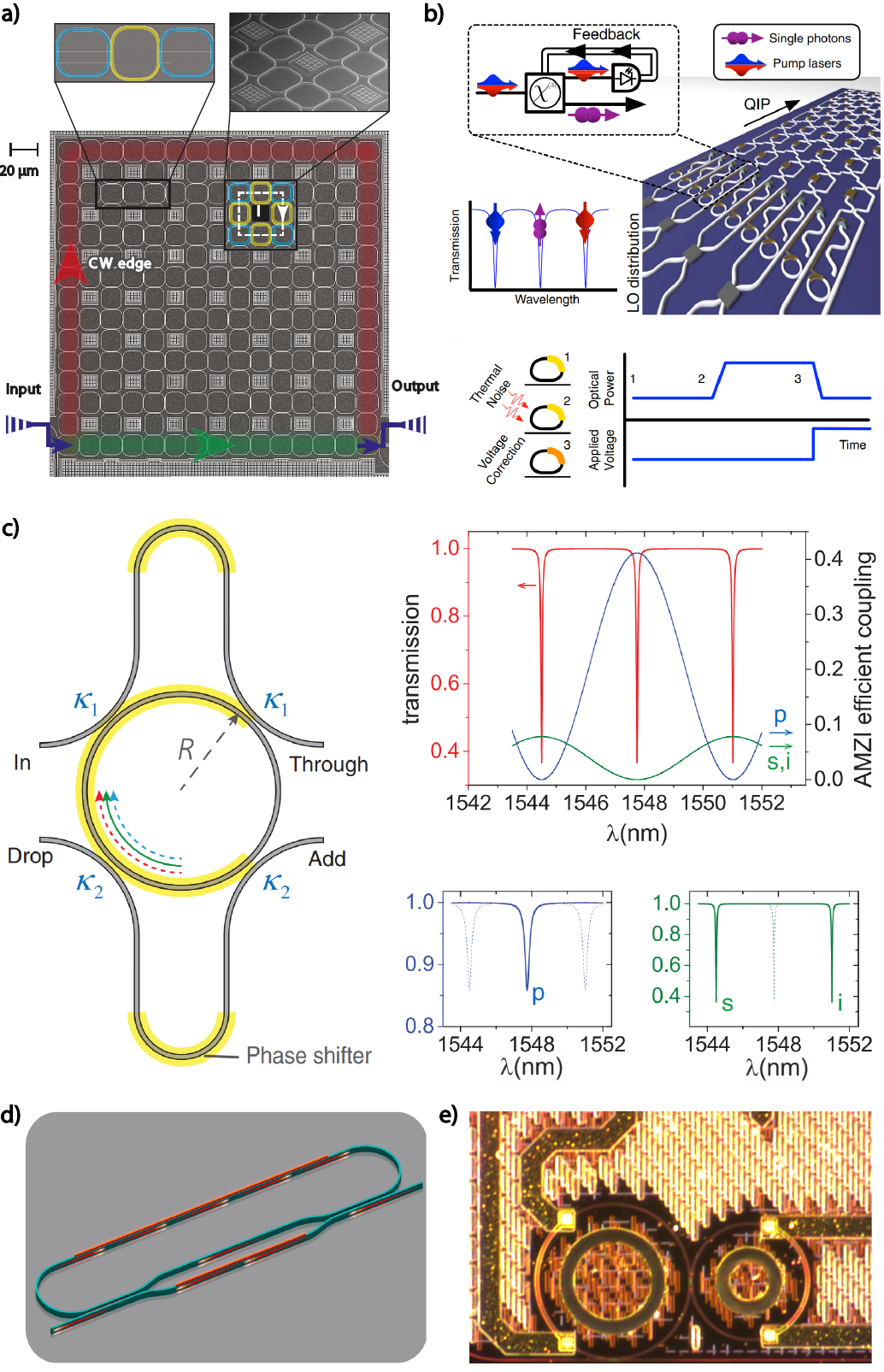}
    \centering
    \caption{Modern ring resonator photon pair sources. (a) A 2D array of ring resonators with topologically protected edge modes is used to generate photon pairs~\cite{mittal2018topological}. (b) PID feedback from a photodiode and controller is a scalable solution to stabilise ring resonances for robustness to environmental noise and cross talk~\cite{carolan2019scalable}. (c) \ch{Asymetric MZI coupling provides tuning of the signal and idler $Q$ factors with respect to the pump~\cite{vernon2017amzimrr}}. A JSA of \ch{spectral purity} of $0.95$ was measured using this structure by a stimulated FWM measurement~\cite{liu2020high}. (d) \ch{A MZI-coupled ring resonator. Here, double-pulse scheme is used to lower the $Q$ factors of the signal and idler resonances with respect to the pump, increasing photon spectral purity, as prescribed in ref.~\cite{christensen2017pump}. A spectral purity of $0.98$ is measured via stimulated FWM~\cite{burridge2020high}}. (e) A dual-ring scheme used to suppress unwanted resonances in a dual-pump scheme for squeezed light generation~\cite{zhang2020single}. \ch{A single ring's regularly spaced resonances preserve energy conservation for FWM to occur between all resonances. Here, the auxiliary ring is engineered to split resonances such that energy conservation is not present for all but the desired FWM process} (Reproduced with permission, \textcopyright~Xanadu)}
    \label{fig:rings}
\end{figure}

One approach to produce pure photons in silicon photonics is with ring resonators~\cite{clemmen2009continuous}.
These resonances act to enhance the SFWM process at specific wavelengths, and can be engineered to produce a more symmetric JSA.
Spectral purity up to $0.93$ theoretically possible with simple ring resonators~\cite{helt2010spontaneous, silverstone2015qubit, vernon2017amzimrr}, though demonstrations were initially limited to $V_\mathrm{HOM} = 0.67$~\cite{faruque2018chip}.
Utilising photon pairs from more than one ring resonator demands accurate process and control, as the shape and positions of the resonances (given by the ring's coupling, loss, and optical path length) are very sensitive to fabrication imperfection---resonances widths are typically around 100 pm.
Though the position of the resonances can be tuned by modulating the phase inside the ring, crosstalk and other environmental factors such as temperature are a issue for scalability (resonances drift by as much as $80$ pmK$^{-1}$~\cite{faruque2018chip}). 
Despite this, high-quality overlap between four ring resonator sources across a large set of on-chip phase configurations has been demonstrated by using an extensive crosstalk characterisation strategy~\cite{llewellyn2020chip}.
Recently, a device was demonstrated that used a photodiode and a PID feedback loop to stabilise resonances achieved stability of less than 1\% of linewidth drift under exacerbated noise (Fig.~\ref{fig:rings}b)~\cite{carolan2019scalable}.  
The approach is scalable, given the photodiode and logic can be implemented locally---either with electronic co-integration, or in an adjacent, or flip-chip-bonded CMOS die (Fig.~\ref{fig:plex}c).

However, purities greater than $0.93$ are required for scaling beyond a few photons. 
In ring resonator sources this can be achieved by shaping the signal and idler resonances with respect to the pump, for example by multi-ring structures~\cite{gentry2016tailoring}, or asymmetric MZI coupling (Fig.~\ref{fig:rings}c)~\cite{vernon2017amzimrr}.
In this technique, the pump resonance linewidth is broadended (lowering the associated quality factor) with respect to the signal and idler resonances.
Recently this scheme was experimentally investigated, with purities of up to $0.95$ and a heralding efficiency of $52$\% ~\cite{liu2020high}, and subsequently implemented on a silicon device generating and manipulating entanglement of two photonic qutrits~\cite{lu2020three}.
\ch{Another approach is to use a split-pulse scheme manipulate the light dynamics inside the ring~\cite{christensen2017pump}.
Here, the first half of the pulse has opposite phase to the first, effectively de-exciting the pump resonance. 
This effectively lowers the pump resonance quality factor, achieving a similar result to~\cite{vernon2017amzimrr}.}
This scheme was recently explored with an MZI coupled ring resonator (Fig.~\ref{fig:rings}d), achieving a maximal spectral purity of $0.98$ measured via stimulated emission tomography~\cite{burridge2020high}, though purity-rate trade off was observed.

Two-dimensional arrays of ring resonators exhibit topologically protected edge states, which may be one route to systematically reduce the effect fabrication-induced disorder on device performance.
Originally considered for delay lines and photon storage, \cite{hafezi2011robust}, topological effects have now been measured using a silicon ring resonator array~\cite{mittal2016measurement} and applied to photon pair production (Fig.~\ref{fig:rings}a)~\cite{mittal2018topological}, though the device's large size and the presence of other resonant modes currently limits applicability.

Recently, a source based on multimode waveguides was demonstrated (Fig.~\ref{fig:sources}b,c,e), with record $V_\mathrm{HOM} = 0.96$\%, spectral purity $P = 0.99$ and a \ch{spectral overlap $0.99$~\cite{paesani2020near} between separate sources}.
In this passive scheme, a pump pulse is split into two waveguides, and one of them undergoes a delay of 13 ps (about $1$ mm). 
Subsequently, a multimode coupler transforms the modes of the two waveguide modes into the TM0 and TM1 modes of a wide crosssection waveguide.
In this waveguide the group velocity of the TM1 mode is greater than the TM0 mode, and so the delayed TM1 mode pulse catches up with and overtakes the TM0 mode pulse.
With gaussian pulses the nonlinear interaction between the pump pulses has a gaussian shape, required for separable photon generation.
The finite (90\%) intrinsic heralding efficiency of the source comes from propagation losses inside the source, which may be improved in the future by sidewall smoothing techniques (see Sec.~\ref{sec:passivecomps}), or shorter pulses.
Importantly, the authors discuss routes for improvement, claiming that heralded HOM interference of greater than 99.9\% should be possible with 4 nm lithographic precision, and truly gaussian pulses.

\subsection {Multiplexing for deterministic photon generation}
\label{sec:determinism}

Photon output probability, $p$, is the leading factor holding back multiphoton devices---in which rates scale as $p^n$---both today, and in the NISQ era ahead. 
 Eventually loss tolerant bounds of around $0.99$ generation and transmission product must be met~\cite{morley2018loss, rudolph2017optimistic}, escaping the exponential curse.
Today, multiphoton experiments in silicon photonics based on SFWM have $0.01 < p < 0.05$.

\begin{figure}[t!]
    \includegraphics[width=.48\textwidth]{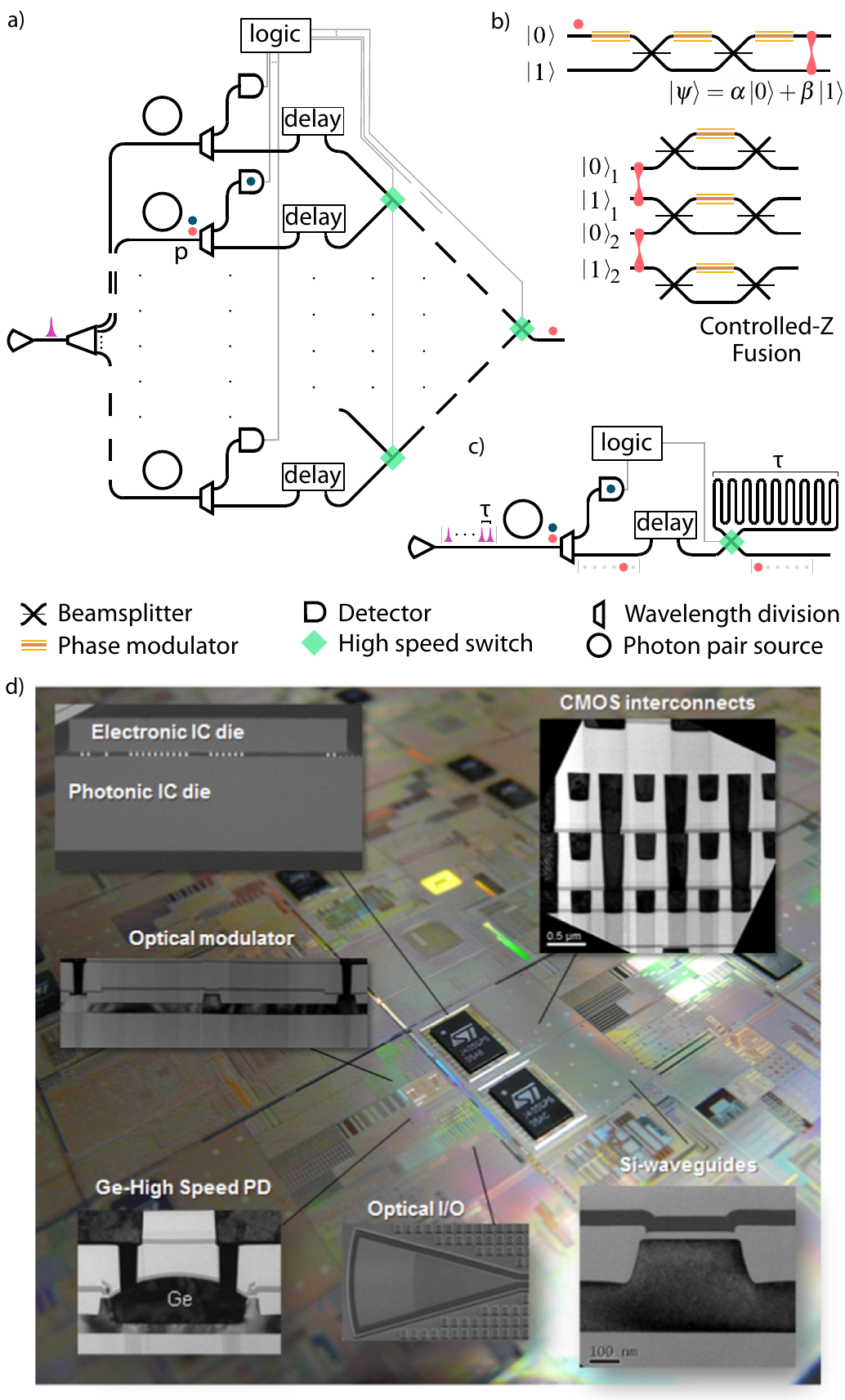}
    \centering
    \caption{Multiplexing and classical logic. (a) Spatial multiplexing of pair photon sources to increase photon probability $p \rightarrow p_m$. The non-degenerate photon pairs are split by wavelength and one of the pair (the herald) is sent to a detector. If a herald photon is detected, it's partner photon enters an optical delay while a logarithmic switch network is activated to route the photon to the source output~\cite{migdall2002tailoring}. (b) A three-phase MZI implements an arbitrary unitary on a single path-encoded qubit, and a reconfigurable two-qubit gate can implement either a controlled-$Z$ or a fusion by postselection~\cite{adcock2018hard}. (c) Temporal multiplexing with a photon storage loop~\cite{pittman2002single, kaneda2019high}. Here, time bins take the place of spatially separate sources as in (a). Using just one optical delay leads to loss scaling linearly---rather than logarithmically---with the number of multiplexed modes. (d) Electronic photonic co-integration via flip chip bonding is an enabling technology for classical silicon photonics. 3D integration techniques such as this are a clear route to reducing the latency of the logic in a multiplexing scheme, and thereby improving efficiencies\cite{thomson2016roadmap}.}
    \label{fig:plex}
\end{figure}

One solution is to multiplex the output of $m$ pair photon sources: using the idler photon to heralded the presence of a signal photon and routing that signal photon via a switching network to a single output (Fig.~\ref{fig:plex}a)~\cite{migdall2002tailoring, pittman2002single}.
Ignoring losses, $p=1$ is asymptotically reached as $m$ is increased.
However, a loss of just 1 dB per switch limits the economy of the technique to just a few multiplexed sources~\cite{bonneau2015effect}.
For this reason, demonstrations of multiplexing have shown modest enhancements of $p$ of around a factor of $E=2$~\cite{mendoza2016active, francis2016all, hoggarth2017resource, kaneda2015time}, including demonstrations using silicon waveguides~\cite{xiong2016active, collins2013integrated}.
Note that sources which are orthogonal by any degree of freedom can be multiplexed, for example, photons in different time bins (Fig.~\ref{fig:plex}c) ~\cite{pittman2002single}.

A recent bulk optical breakthrough demonstrates enhancements up to $E = 28$ for low pump powers (low base $p$), and $E=10$ at high $p$ by temporal multiplexing via a ultra-low-loss photon storage loop. 
The experiments peak output probability was $p_m = 0.67$ (to optical fiber)~\cite{kaneda2019high}.
Other techniques, like multiplexing in frequency modes~\cite{joshi2018frequency}, offer density, but require frequency conversion, which today is costly and inefficient.
These may be combined to multiplicatively increase the number of multiplexed modes.

To multiplex pair photon sources in silicon photonics, a fast, low-loss switch is required, such as the solutions based on hybrid integration of TFLN and BTO (see Sec.~\ref{sec: modulators}).
These modulators achieve modulation speeds of tens of GHz and have losses limited by propagation losses (as low as $2.7$ dB/m).
A time loop multiplexing device operating a $10$ GHz requires a delay line of just 0.8 cm of silicon, enabling a switch-loop transmission of $-0.05$ dB (98.8\%), equalling the loss of the storage loop of ref.~\cite{kaneda2019high}.
Meanwhile, scaling up to more than $50$ GHz may be possible~\cite{He19NPLNmod, Abel19BTO}.

The dominant loss in this technology will due to the detection-to-switch-activation latency, forming a barrier to multiplexed photon probability.
Ref.~\cite{kaneda2019high} demonstrated a switching latency of 100 ns (via an FPGA), which would require prohibitive waveguide lengths.
Routing photons off chip to low-loss optical fiber is one option, however with state-of-the-art grating couplers this represents a maximum heralding efficiency of $0.85$, even with no other losses considered.
Still, the potential improvements from today's single digit efficiencies is a large one.

In the future, ultra-low latencies must be achieved by close integration of silicon logic, for example by directly wirebonding, or flip-chip bonding a CMOS logic circuit (Fig.~\ref{fig:plex}d).
With these techniques electronic latencies of $1$ ns or lower may be achieved.
$1$ ns corresponds to around $8$ cm of silicon waveguide delay line, which could have a transmission of $-0.2$ dB (95\%) with state-of the art silicon waveguides.
Delay and storage losses could be reduced further by use of silica or silicon nitride hybrid integration, which potentially offer up to an order of magnitude improvement to propagation loss (see Sec.~\ref{sec:passivecomps}).
Low-loss optics in a monolithic electronic-photonic platform will enable the ultimate in low-latency and scaling, with recent progress in classical silicon photonics showing great promise~\cite{sun2015single, atabaki2018integrating, chung2018monolithically}.
Today, post-photonic processing steps typically degrade optical performance, making low-loss operation a challenge.
Meanwhile, photon pairs have been generated in the silicon layer of a CMOS platform~\cite{gentry2015quantum}.

The same feedforward capability that enables multiplexing also enables the two-qubit KLM gate~\cite{knill2001scheme}, as well as adaptive measurement, a key ingredient of measurement-based computing.
Meanwhile, hierarchical multiplexing of photons and entanglement generation enables the linear optical quantum computer architecture of ref.~\cite{gimeno2015three} (see Section~\ref{sec:processing}.)

\subsection {Engineering nonlinear sources}
\label{sec:problems}

Though it provides a powerful and easy-to-use route for single photon production, silicon's $\chi^{(3)}$ non linearity has some limitations.
At 1550 nm, the generation of free electrons by two-photon absorption (TPA) and subsequent free-carrier absorbtion places limits on the heralding efficiency of SFWM-based photon pair sources~\cite{husko2013multi}, for example a photon generation rate of $p=0.01$ implies an intrinsic heralding efficiency of $0.96$~\cite{husko2013multi, silverstone2016silicon}.
To reach loss tolerant thresholds around $0.99$, $p$ must be reduced by an order of magnitude to $p=0.001$.
While low $p$ also reduces multiphoton noise (also necessary for fault tolerance), the number of multiplexed sources needed to produce a deterministic single photon rapidly grows to infeasible levels~\cite{bonneau2015effect}.
One solution is operate where two photons do not have enough energy to excite and electron across the bandgap of silicon, that is, at wavelengths of $2.1$ $\upmu$m or greater, in the mid infrared.
Recently quantum interference was demonstrated in the mid-infrared in specially engineered silicon waveguides, showing the potential to banish TPA\footnote{Interestingly, this was the first quantum interference in the mid-infrared across optics~\cite{rosenfeld2019first}.} (Fig.~\ref{fig:sources}a)~\cite{rosenfeld2019mid}.

TPA is not the only deleterious nonlinear effect experienced as the pump and single photon fields co-propagate.
Self- and cross-phase modulation of the pump and single photons, as well as dispersion and absorption from free carrier all play a role.
Modelling these effects is nontrivial, especially in ring resonators where large field enhancements produce complex dynamics~\cite{ma2020prospects}.
Here, nonlinear, thermal and carrier alter the ring's resonant conditions, leading to distortions of the resonances and subsequently of the generated photons' JSA.
A recent theoretical investigation in to these effects in straight waveguide sources found that these effects do not play a large role in the spectral purity of the emitted single photons---spectral purity can even be increased by nonlinear pump broadening~\cite{sinclair2016effect}.
Meanwhile, ref.~\cite{koefoed2019complete} details an efficient, split-step Fourier method for simulation of four-wave mixing, with a prescription for including silicon-specific effects such as TPA.

\ch{
\subsection {Hybrid integration of single photon emitters}
\label{sec:solid}

Semiconductor quantum dot single photon sources~\cite{senellart2017high} show great promise for future integrated quantum photonic technology.
High-performance hybrid integration of these systems---typically III-V semiconductors---with a mature integrated photonics platform is a long-standing goal of the field.
These sources have been under development for over two decades, and in recent years have attained key performance metrics.
For example, advances in cavity design and pulse engineering have boosted 
indistinguishability to over $0.99$ and transmission to single mode fiber to up to $0.15$~\cite{somaschi2016near}.
Meanwhile, the emission mode coupling factor, $\beta$, can be as high as $0.98$~\cite{arcari2014near} to suspended AlGaAs waveguide structures, or $0.65$ in free space~\cite{somaschi2016near}, giving access to the regime of strong coupling between light and matter.
Other innovations, such as wavelength tunability by electronic integration or by induced strain have also been utilised to produce HOM interference between independent quantum emitters~\cite{patel2010two, flagg2010interference}---a crucial capability for scalability.

The most mature quantum dots photon sources typically emit around 900 nm, however, silicon is opaque below around 1100 nm.
Therefore, other solutions must be found to couple with silicon waveguides, and furthermore to be compatible with terrestrial quantum communications systems, which rely on the high-transmission window of optical fibers around 1550 nm. 
Quantum dots at telecommunications wavelengths have been demonstrated~\cite{birowosuto2012fast, haffouz2018bright, muller2018quantum}, though the technology is less mature, with distinguishability and brightness metrics lagging behind their near-infrared counterparts.
However, recent demonstrations of integration with silicon nitride waveguides---which is transparent across into the visible band---show promise.
Other techniques, such as quantum frequency conversion, can be used to reach the 1550 nm band: two-photon interference between initially distinguishable, approximately 900 nm photons emitted from independent quantum dots were recently interfered after frequency conversion to the telecommunications band~\cite{weber2019two}.

There has also bee much progress in integrating quantum dots with waveguide circuits.
In one demonstration nanopillar cavities containing quantum dots were be placed on a substrate using a micromanipulator, and silicon nitride waveguides are subsequently lithographically defined to encapsulate it~\cite{zadeh2016deterministic, elshaari2017chip}.
The vertically pumped dot achieved a $g^{(2)}(0) = 0.07$, and has a good coupling to the waveguide mode of $\beta = 0.24$, though simulations imply coupling of up to $0.9$ is possible.
Due to its change in environment, the emission of the source shifts by several linewidths, which may be a hurdle to operating multiple indistinguishable sources.
A followup~\cite{elshaari2018strain} result demonstrates piezo-electric strain tuning of the sources as a possible route to wavelength control and thereby scalability.
A waveguide quantum dot source with photonic crystal back-reflector has been used to demonstrate similar functionality on silicon waveguides at around 1300 and 1410 nm~\cite{kim2017hybrid}, including an on-chip splitter used to measure the photon's  $g^{(2)}(0) = 0.33$ and a coupling factor of $\beta = 0.32$.
A similar result using silicon waveguides is achieved in refs.~\cite{katsumi2019quantum, katsumi2018transfer}, with emission at 1150 nm and a  $g^{(2)}(0) = 0.3$

Using an advanced lithographic process, ref.~\cite{davanco2017heterogeneous} demonstrates a tapered waveguide of GaAs directly on top of a silicon nitride waveguide.
Here, pump light in the silicon nitride waveguide adiabatically couples to the GaAs waveguide, which hosts a selection of InAs quantum dots.
Resonant enhancement is provided by either a photonic crystal cavity in the GaAs waveguide, or a GaAs ring resonator, achieving a coupling of $\beta = 0.20$ and a  $g^{(2)}(0) = 0.13$ for the photonic crystal.
Native integration with low-loss silicon nitride waveguides provides one pathway scalability---if the optical properties (emission spectra, purity) can be controlled (perhaps via improvements in processing or in-built strain or electrical tunability).

Meanwhile, various other quantum emitters show promise for waveguide integration.
For example, carbon nanotube photon sources~\cite{khasminskaya2016fully}, or quantum dots in 2D materials~\cite{peyskens2019integration}.
Ref.~\cite{wan2020large} describes the coupling of of 16 arrays of 8 diamond vacancy single photon sources with AlN waveguides, showing $0.05 <g^{(2)}(0) <0.5$ and 50 GHz of wavelength tunability across the 128 samples.
Here, tapered diamond waveguides containing silicon and germanium vacancy centers are vertically pumped to produce single photons into the coupled waveguide, achieving $0.01 < g^{(2)}(0) < 0.5$ across 128 samples.
Though these centers emit in the visible band and couple to AlN waveguides, the work demonstrates the maturity of micromanipulation techniques for placing structured arrays, which may be applied to other systems, such as 1550 nm silicon carbide vacancy sources or to silicon nitride waveguides.

}

\section{Quantum information processing}
\label{sec:processing}
The last two years has seen an explosion of progress in large-scale silicon quantum photonic experiments.
In this section we will examine these devices and their applications.

\subsection{Encoding quantum information in silicon waveguides}

Any two distinct states of a single photon can comprise a qubit---a good qubit is one with a long lifetime and which can be controlled with high fidelity and speed.
The dominant qubit encoding used for quantum information processing in silicon photonics is path encoding in two waveguides.
Here, the $\ket{0}$ and $\ket{1}$ states of the qubit correspond to a single photon occupying the first and second waveguide respectively, and is generalised to qudits by using $d$ waveguides.
Local unitary operations on path qubits are implemented by using a three-phase MZI (Fig.~\ref{fig:plex}b) which have been demonstrated to have less than -66 dB of intrinsic nois~\cite{wilkes201660, harris2017quantum}.
Meanwhile, the generalisation to for $d$ dimensional is provided by a universal interferometer, for example the \emph{Clements scheme}~\cite{clements2016optimal}.
With no demonstrated single-photon nonlinearity on the platform, two-qubit gates are currently limited and rely on postselection, or on heralding (consuming photons).
Other qubit encodings, such as polarisation \cite{takesue2008generation, matsuda2012monolithically, ma2016silicon}, frequency, higher-order modes~\cite{feng2019chip}, time bins~\cite{sibson2017chip} have been explored, but are less common for information processing.

\subsection{Large-scale circuits and Multiphoton capability}

\begin{figure}[t!]
    \includegraphics[width=0.48\textwidth]{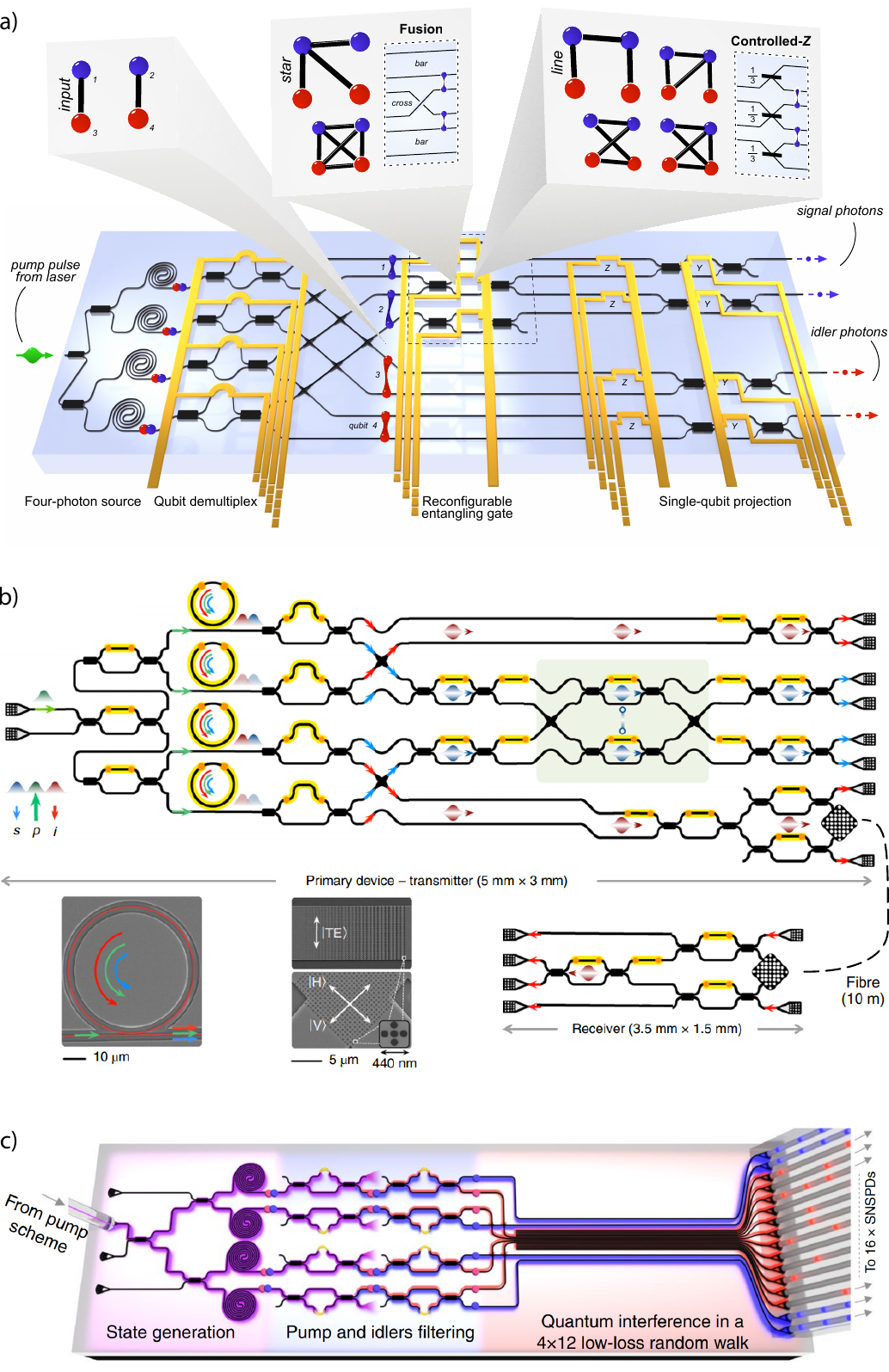}
    \centering
    \caption{Silicon quantum photonic circuits with multiphoton capability. (a) A four-photon device generating every type of four-qubit graph state. Here, fidelities were limited to $0.68$ and $0.78$ by $0.80$ visibility HOM interference and low photon-flux due the use of standard, non-optimised components~\cite{adcock2019programmable}. (b) Chip-to-chip quantum teleportation via generation of a four-photon entangled resource. A ring resonator source is shown in the inset, along side a 2D grating coupler for path-to-polarisation conversion~\cite{llewellyn2020chip}. (c) A silicon chip generating up to eight degenerate photons for boson-sampling with a fixed unitary implemted with a MMI~\cite{paesani2018generation}.}
    \label{fig:multiph}
\end{figure}

\begin{figure*}[ht!]
    \includegraphics[width=1\textwidth]{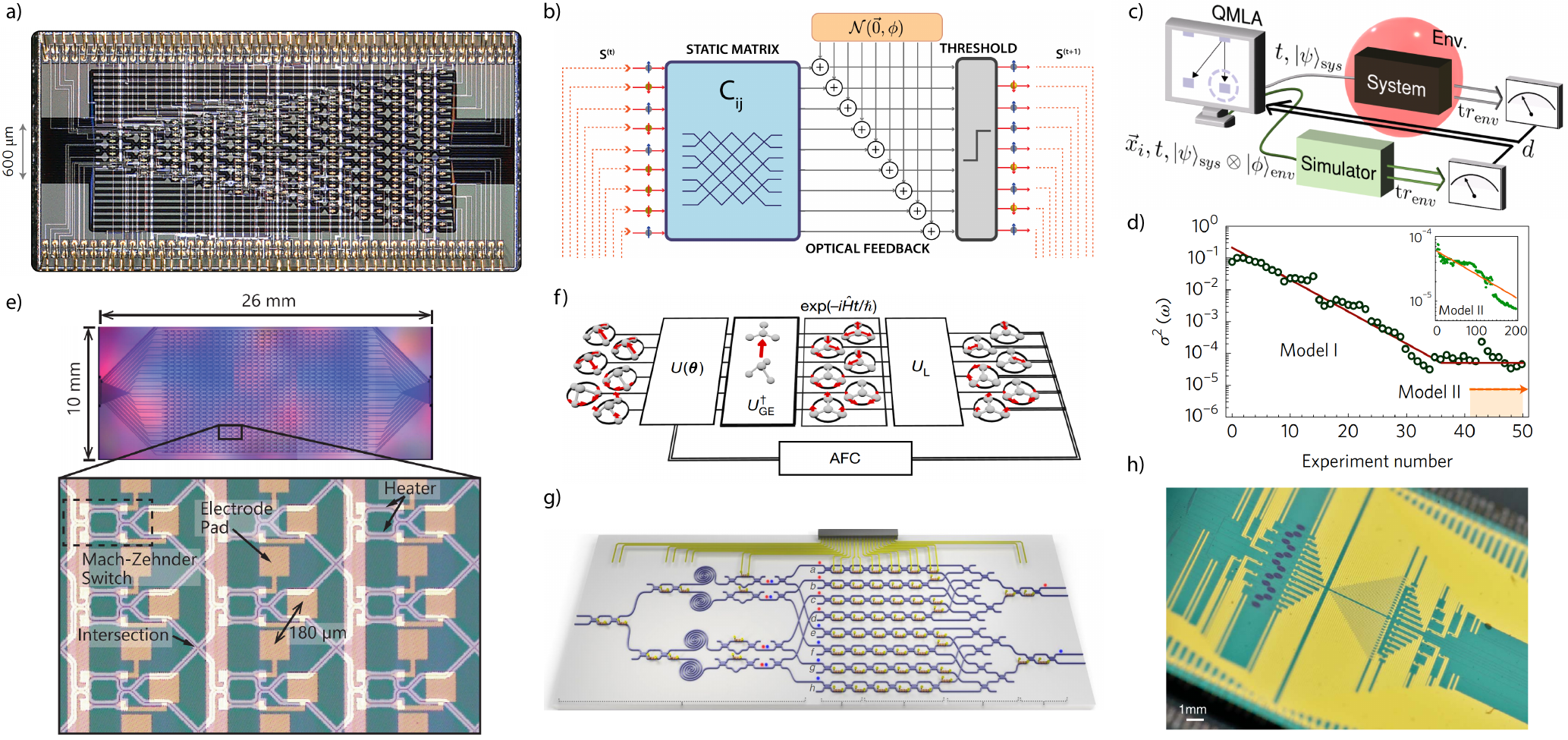}
    \centering
\caption{Large-scale silicon quantum photonics and machine learning. (a) A large scale universal photonic chip. \ch{Here, a $3\times26$ mesh of connected MZIs allow the implementation of arbitrary linear optical unitary transformations. The top and bottom of the image are device's 194 wirebonds, demonstrating complex package required for operating large scale circuits.} This universal linear optical circuit was used as a testbed for numerous experiments such as quantum transport simulations~\cite{harris2017quantum} and interactions in a coherent Ising machine~\cite{prabhu2020accelerating}, whose layout is shown in (b). Here, the chip implements the matrix $C_{ij}$. (c) Quantum Hamiltonian learning: A classical optimiser learns the Hamiltonian of an unknown quantum system with no a priori information via a quantum simulator~\cite{gentile2020learning}, \textcopyright~The Authors. (d) Optimising Hamiltonian parameters with a silicon photonic quantum co-processor~\cite{wang2017experimental}. (e) Microraph of a $32\times 32$ silicon photonic switch with an average loss of $10.8$ dB. The device's 1024 phase modulators are addressed via a vertical interposer~\cite{Suzuki2019low}. (f) Algorithm for modelling vibronic spectra of molecules with a universal optical circuit~\cite{sparrow2018simulating}. (g) Schematic for universal quantum photonic processor which computes 2-qubit quantum logic via the Cartan decomposition~\cite{qiang2018large} \ch{This decomposition allows any two-qubit unitary ($\textrm{SU}(4)$ martrix) can be written as a sum of four local unitaries \cite{zhou2011adding}. The device utilises parallel MZIs to implement local unitaries which are later combined.}. (h) Photograph of A large-scale silicon quantum photonic device featuring $16$ SFWM photon pair sources and over $500$ passive components~\cite{wang2018multidimensional}. \ch{This device was used to generate a bell state of $d=15$ qudits, $\sum_{i}^{15}{\ket{ii}}/\sqrt{15}$, and served as a testbed for demonstrating nonlocality and steering in high dimensions. }}
    \label{fig:ML}
\end{figure*}

The attraction of silicon photonics comes from its scalability.
Once a technique for generating a pair of qubits is established~\cite{silverstone2014chip, silverstone2015qubit}, scaling the number components on a design is relatively simple (given loss-budgeting).
Commercially available V-groove fiber arrays and multichannel analogue electronics have enabled the rapid emergence large-scale quantum photonics.
For example, the demonstration of 15-dimensional entanglement using a device with simultaneously operated 15 photon pair sources\footnote{Arrays of 128 photon pair sources of $0.9$ indistinguishability have also been demonstrated in silica\cite{ren2020128}, though at most three have been demonstrated simultaneously~\cite{spring2017chip}.} and over 500 passive components (Fig.~\ref{fig:ML}h)~\cite{wang2018multidimensional}.
Similarly, a universal two-qubit processor was implemented via a postselected photonic decomposition of $SU(4)$ (Fig.~\ref{fig:ML}g)~\cite{qiang2018large}.
This processor featured $0.92$ fidelity two-qubit gates and was used to demonstrate NISQ algorithms, such as quantum approximate optimization algorithm, and an efficient simulation of directed quantum walks---albeit with two qubits.
This improves on a smaller-scale circuit based on a postselected controlled-$Z$ gate (Fig.~\ref{fig:plex}b) implemented between a pair of degenerate photons which achieved state fidelities of $0.8$ and above~\cite{santagati2017silicon}.

Though off-chip quantum interference between silicon sources was shown in the early days of the field~\cite{harada2011indistinguishable, zhang2016interfering}, on-chip multiphoton capability---that is, the generation, interference, entanglement and control of multiphoton states in silicon waveguides---is the first step to scaling up.
On-chip heralded HOM interference was demonstrated for the first time between photons generated in two independent ring resonator sources~\cite{faruque2018chip} in 2018, achieving a visibility of $V_{\mathrm{HOM}} = 0.67$.
The success of this experiment relied on careful tuning and crosstalk mitigation in the photon-generating ring resonances (see Sec.~\ref{sec:source}).
Subsequent demonstrations have implemented photonic circuitry to entangle up to four photons in path-encoded qubits.
For example, a reconfigurable postselected entangling gate~\cite{adcock2018hard} was used to programmably generate both cluster and GHZ resource graph states (Fig.~\ref{fig:multiph}a)~\cite{adcock2019programmable}---resources for measurement-based quantum information protocols.
The generation of graph states from nonlinear sources and postselected entangling gates was studied in ref.~\cite{adcock2018hard}, revealing that most graph states are not accessible, echoing the call for the development of deterministic technology.
A similar experiment produced GHZ entanglement as a resource for quantum teleportation between remote devices\cite{llewellyn2020chip} using ring resonator sources (Fig.~\ref{fig:multiph}b).
A four-photon $\ket{\Phi^+}\otimes\ket{\Phi^+}$ state was also recently generated via a silicon waveguide \cite{zhang2019generation}.

In any quantum photonic device, photon throughput determines the eventual data rate, while the computational space accessed by $n$ qudits of dimension $d$ is $d^n$. 
Doubling $d$ (equivalent to adding a qudit) roughly results in circuit transmission $\eta \rightarrow \eta^2$ for a $d$-dimensional local unitary \cite{clements2016optimal}, while adding a photon source with efficiency $p$ incurs a cost $\eta p$.
Today, with $p < 0.2$, increasing $d$ is an attractive route to larger state spaces---though optimum values of $n$ and $d$ depend on the exact photonic design and its loss.

\subsection{The NISQ era and beyond}

Quantum simulations and machine learning-enabled quantum information processing are poised to be major applications of quantum technology especially in the NISQ era~\cite{biamonte2017quantum, adcock2015advances, georgescu2014quantum}.
Many of these applications have natural photonic implementations, with quantum optical neural networks~\cite{steinbrecher2019quantum}, variational (hyrbid quantum-classical) algorithms \cite{peruzzo2014variational}, optimization algorithms~\cite{qiang2018large} and coherent Ising machines~\cite{yamamoto2017coherent} being prominent examples.
Recently a silicon photonic circuit featuring controlled unitary was used to implement a Bayesian extension of the quantum phase estimation algorithm, enabling increased robustness and phase precision~\cite{paesani2017experimental}, and was used to simulate a hydrogen molecule.
These methods were adapted into a new variational\footnote{In variational quantum algorithms, a set of parameterised quantum gates (i.e.~those reconfigurable in the photonic circuit) is subject to classical optimisation and repeated to find a solution (for example, the ground state of a Hamiltonian).} protocol to find eigenstates of a Hamiltonian, which the authors name \emph{WAVES}~\cite{santagati2018witnessing}---though good ansatze are required for efficient operation.
A related problem was also addressed, where the Hamiltonian of a nitrogen vacancy (NV) center in diamond was learned~\cite{wang2017experimental}.
Here, single photon data from the NV centre was compared to data from the chip, which acts as a quantum co-processor, simulating the NV system.
This technique exhibited the ability to detect that new terms may be needed in the Hamiltonian (Fig.~\ref{fig:ML}d), and has since been extended to learn a Hamiltonian with no prior form (Fig.~\ref{fig:ML}c)~\cite{gentile2020learning}---though here the quantum co-processor was simulated.

Recently, a single $3\times 26$ universal silicon photonic processor (Fig.~\ref{fig:ML}a) has been used to model quantum particle transport experiment by single photon quantum walk (with laser light)~\cite{harris2017quantum} as well as in a counterfactual communication experiment with single photons~\cite{calafell2019trace}.
The same device was also used as a proof-of-principle recurrent Ising machine experiment~\cite{prabhu2020accelerating}---a promising optical analogue to quantum annealing in the solid state (Fig.~\ref{fig:ML}b)~\cite{johnson2011quantum}.
Here, all-to-all connections provided by the universal unitary enable the encoding of arbitrary four-spin Ising problems which are inefficient to encode in quantum annealers~\cite{hamerly2019experimental}. 
Though a 100 spin device was recently demonstrated using time-bin encoding using optical fiber~\cite{mcmahon2016fully}, future silicon photonic implementations should enable large speed ups with parallel processing.
This circuit was also used to study an advanced calibration method where physical imperfections (for example non 50:50 beamsplitter reflectivities) are `calibrated out' using a nonlinear optimisation, resulting in 10 dB improvements to measured process fidelities~\cite{mower2015high}.
Finally, the same device was used to learn a the `black box' unitary implemented in a boson sampling experiment via a variational method, given known input and output samples~\cite{carolan2020variational}.

Outside of qubits, boson sampling\footnote{see ref.~\cite{BosonSamplingReview} for a comprehensive review of boson sampling.} is a computational task that is prominent in quantum information literature as a route to demonstrations of quantum advantage, and has a variety of computational applications \cite{sparrow2018simulating, bromley2020applications}.
Sampling from the distribution of measured output positions produced by bosons traversing a unitary matrix (linear optical network) is known to be of exponentially complexity~\cite{aaronson2011computational, clifford2018classical}, whereas lossless physical implementations using photons are trivially subject to linear scaling.
Gaussian boson sampling is a promising route to applications in the NISQ era, enabling a host of graph-based computations as well as the computation of molecular vibronic spectra (Fig.~\ref{fig:ML}f)~\cite{bromley2020applications, sparrow2018simulating}.
Recently, states of up to eight degenerate photons were generated in a recent scattershot and Gaussian boson sampling experiment, featuring a fixed optical unitary implemented with a fixed multimode interference region (Fig.~\ref{fig:multiph}c)~\cite{paesani2018generation}.
Meanwhile, a recent bulk-optical demonstration of boson sampling demonstrated 20 photons (up to 14 simultaneously detected) in a 60 mode fixed interferometer using photons demulitplexed from a quantum dot source, stepping towards the demonstration of quanutm advantages~\cite{wang2019boson}. 
To compete, a silicon photonic implementation must drastically reduce loss, for example by inclusion of sidewall smoothing techniques in advanced fabrication processes (See Sec.~\ref{sec:passivecomps} and Table~\ref{tab:losses}).
Meanwhile, a $32\times32$ silicon photonics switch with average loss of $10.8$ dB was recently demonstrated (Fig.~\ref{fig:ML}e) using a vertically bump-bonded interposer accessing the device's 1024 phase shifters~\cite{Suzuki2019low}, though this device---a result of photonic interconnect research---lacks phase control required for implementation of arbitrary optical unitaries~\cite{clements2016optimal}, its scale and low-loss is enabled by techniques which are transferable to silicon quantum circuits.

\begin{figure}[t!]
    \includegraphics[width=0.48\textwidth]{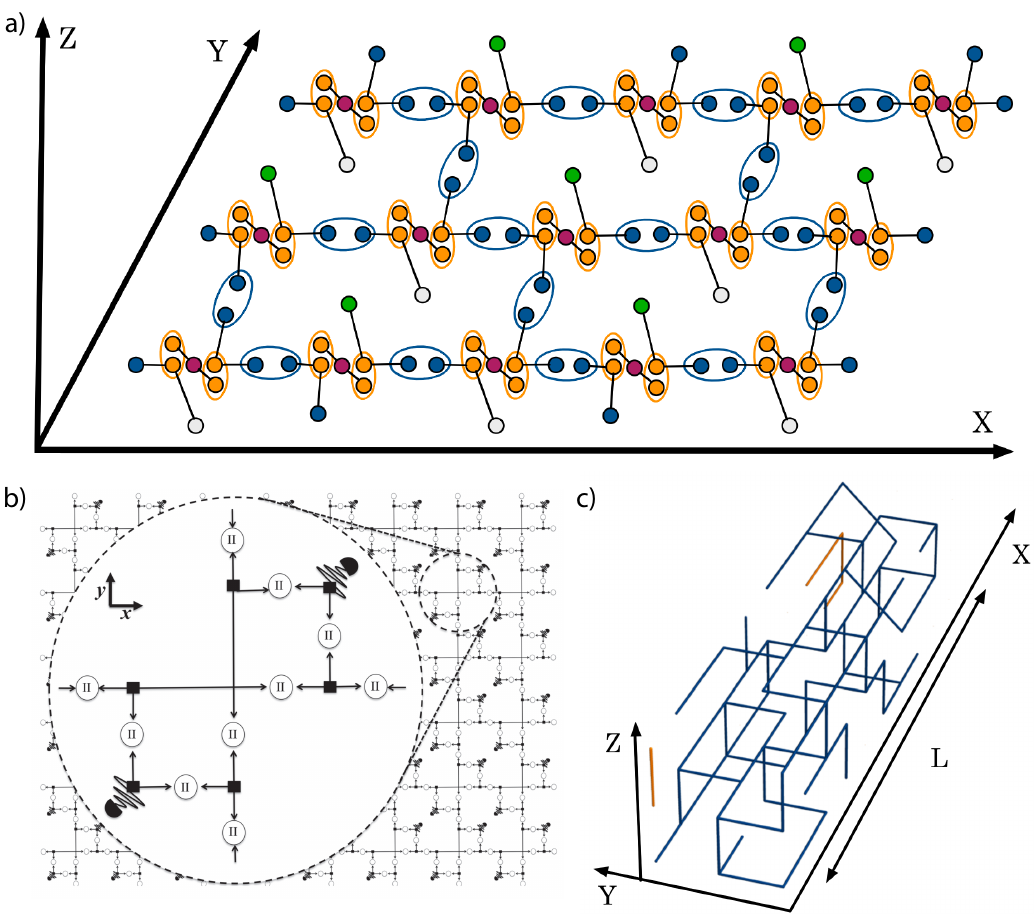}
    \centering
    \caption{Ballsitic linear optical quantum computing architecture. (a) A 3D universal lattice resource state is produced from 3-photon GHZ states via probabilistic fusion operations. Shown here is a single plane of the 3D state, with green qubits linking through the Z direction to white qubits.~\cite{gimeno2015three}. (b) A 2D unit tiling implementing (a)~\cite{rudolph2017optimistic}. (c) The result: an imperfect lattice is produced, where logical qubits are defined on percolated paths through the lattice. Here, the X direction is equivalent to photon time-of-flight.}
    \label{fig:loqc}
\end{figure}

Silicon quantum photonics is also leading candidate platform for linear optical quantum computation, with recent theoretical results closing the gap between today's devices and large-scale, fault-tolerant quantum computers.
The leading architecture is based on the deterministic production of percolated resource states for measurement-based quantum computing via hierarchical multiplexing of single photon sources and entanglement generation~\cite{gimeno2015three}.
When more than one layer of multiplexing is used, a theory of `relative multiplexing' has been developed to provide the optimal scheduling of resources across the whole device~\cite{gimeno2017relative}.
Interesting, the number of components that each photon travels through in this architecture is independent of the total device size (Fig.~\ref{fig:loqc}b)---a key requirement for scalability (cf.~$n$-photon boson sampling which requires each photon to traverse $O(n)$ components, resulting in exponential losses $O(\eta^n)$).
Probabilistic gates are used to fuse deterministically mulitplexed three-qubit GHZ states (Fig.~\ref{fig:loqc}a,c), resulting in a lattice resource state with randomised connections.
In the limit of infinite lattice size, results from percolation theory ensure that logical, error-correctable qubits exist throughout.
Recently, this architecture has been simulated at scale and has been shown to be robust at finite sizes~\cite{morley2017physical}, and meanwhile tolerance to photon loss in resource states has been demonstrated~\cite{morley2018loss}.
Though we observe great progress in the last five years, the challenges of building such a device~\cite{rudolph2017optimistic}---which we summarise in Sec.~\ref{sec:scaling}---remain daunting.

\section{Quantum communication}
\label{sec:comms}
Quantum communication is the faithful transfer of generic quantum states between remote locations~\cite{gisin2007quantum}.
Quantum communications use a variety of qubit encodings, with time-bin, polarisation and spatial modes utilised, depending on application.
During the last five years, there have been many achievements in this area, including entangled distribution over 1200 km~\cite{yin2017satellite}, long distance quantum key distribution~\cite{chen2020sending}, underwater quantum communication~\cite{hufnagel2020underwater, tarantino2018feasibility} and large-scale quantum and classical multiplexing~\cite{bacco2019boosting, eriksson2019wavelength}.
All of these demonstrations used bulk-optical components, which limit applicability---integrated photonics instead offer a route to cost-effective QKD, deployed at scale.

\subsection{Quantum key distribution}
\label{sec:qkd}

\begin{figure*}[ht!]
    \includegraphics[width=1\textwidth]{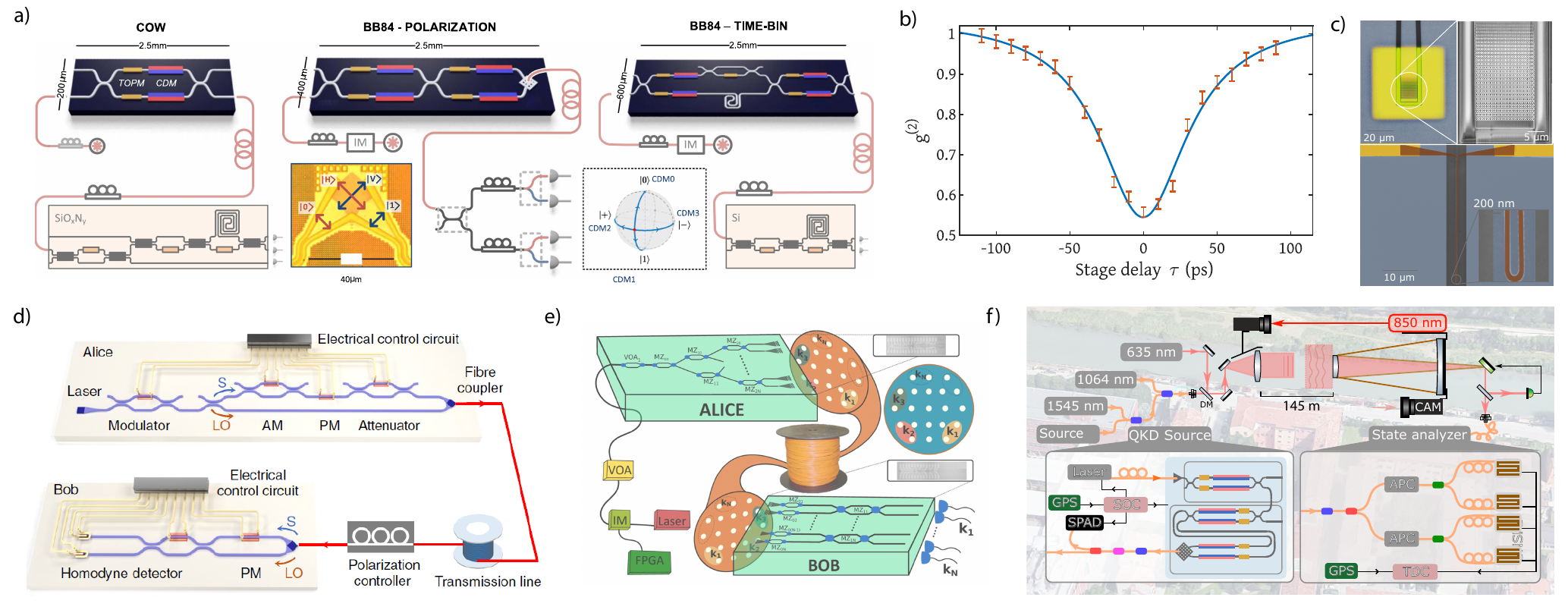}
    \centering
    \caption{Silicon photonics for quantum communications. (a) Silicon photonic chips implementing a range of quantum key distribution protocols~\cite{sibson2017integrated}. (b) \ch{HOM interference between weak coherent pulses---a necessary feature of MDI QKD---generated via hetero-integration of III-IV InP on silicon photonics~\cite{agnesi2019hong}. When interfering coherent light, the maximum achievable visibility is $0.5$---this result ($0.46$) is within theoretical bounds for efficient MDI QKD.} (c) Integrated relay server for MDI QKD, featuring SNSPDs patterned on an a commercial silicon photonic platform (bottom) featuring photonic crystal grating couplers (top) \cite{Zhang19arxivChip}. (d) Experimental setup of a continuous-variable chip-to-chip QKD protocol~\cite{zhang2019integrated}, \textcopyright~The Authors. (e) Space division multiplexing with silicon integrated photonics~\cite{bacco2017space}. (f) Daylight free-space quantum key distribution using silicon photonic circuits~\cite{avesani2019full}, \textcopyright~The Authors.}
    \label{fig:QC}
\end{figure*}

Quantum key distribution (QKD) uses measurement-induced collapse of superposition states to detect eavesdroppers on a quantum communications link, enabling a secure cryptographic keys to be generated between two remote locations. 
To be compatible with telecommunications technology and to minimise loss, operation should be at 1550 nm, with secret key generation rate and link distance desirable figures of merit.
Today, in the absence of quantum repeaters~\cite{briegel1998quantum}, link distance is limited by channel loss.
Recently there have been multiple demonstrations using integrated silicon optics with weak coherent states~\cite{ma2016silicon, sibson2017integrated, avesani2019full} (Figures \ref{fig:QC}a,d,e,f).
In particular, focusing on a discrete variables approach, ref.~\cite{sibson2017integrated} achieves chip-to-chip quantum key distribution using silicon photonics at 1550 nm.
In this work various quantum key distribution protocols were demonstrated (COW, BB84-polarisation encoding and BB84-time-bin encoding) by exploiting 10 GHz bandwidth carrier-depletion modulation.
These were combined with DC-operated thermo-optic modulators to minimise phase-dependent loss errors.
In the same direction, two field-trial experiments, one over free-space in daylight conditions (Fig.~\ref{fig:QC}f)~\cite{avesani2019full} and one in fiber~\cite{bunandar2018metropolitan}, have recently been realised using integrated photonics employing weak coherent states at telecommunications wavelengths.
Here, the use of silicon photonics enabled the experiments high secret key rates and increased the stability of the system, particularly by correcting for polarisation drift.
One recent innovation comes from a polarisation-based QKD transceiver device, where the same circuit traversed in different directions gives provides transmitter or receiver functionality \cite{cai2017silicon}.

Increasing secret key rates is crucial for developing quantum communications networks.
Taking inspiration from classical telecommunication technology, secret key rate can be expanded by multiplexing, for example space division multiplexing was recently demonstrated by use of a multicore fiber (Fig.~\ref{fig:QC}e)~\cite{bacco2017space}.
In this experiment, keys were generated by a parallel silicon photonics transmitters, and coupled to a multicore fiber by a circular grating coupler array, with a similar silicon chip featuring parallel receiver used to decode the stream of qubits. 
High-dimensional quantum key distribution has been recently demonstrated using a technologically similar approach: four-dimensional quantum states were transmitted over a few meters of multicore fiber~\cite{ding2017high}, where different fiber cores corresponded to the different basis states of a photonic qudit.
More recently, this transmission distance was extended to kilometer length scales using phase-locked loops~\cite{canas2017high, da2019stable}.
SNSPDs are among the most costly and least portable resource in quantum photonics---their availability is often a bottleneck in quantum photonics research. 
Time multiplexing detectors between multiple users alleviates this, and was recently demonstrated in a QKD setting with a silicon chip~\cite{kong2020photonic}.

Thus far we have discussed discrete variable QKD, however, other qubit formats can also be used, for example continuous variable (CV) QKD has recently gained attention~\cite{pirandola2019advances, xu2019quantum}. 
This format is preferred for short links ($< 100$ km), due to limited signal-to-noise ratio at long distance. 
Silicon quantum photonics has application here, too, with scalable on-chip homodyne detection, based on high-quantum efficiency photodiodes~\cite{raffaelli2018homodyne}.
Recently, a silicon chip has been employed to generate a gaussian modulation scheme reaching an secret key rate of 0.14 kbps over 100 km of fiber (Fig.~\ref{fig:QC}d)~\cite{zhang2019integrated}.
A local oscillator and the quantum signal---which have orthogonal polarisation---were coupled together into an optical fiber before demultiplexing and measurement through homodyne detection on the device.
Recent theoretical advances in CV QKD have proved a the same level of security as in the discrete variable approach~\cite{Ziebell15, pirandola2019advances}, setting the stage for future developments on silicon and otherwise.

Measurement-device-independent (MDI) protocols allow the creation of a quantum key without relying on a trusted receiver, improving security.
Toward this goal, $0.46$ visibility Hong-Ou-Mandel interference---crucial for MDI protocols and limited to $0.5$ for coherent states~\cite{wei2019high, semenenko2020chip}---was recently demonstrated (Fig.~\ref{fig:QC}b)~\cite{agnesi2019hong}) between two integrated distributed feedback lasers on discrete silicon chips---simultaneously with a result using indium phosphide devices~\cite{semenenko2019interference}.
Recently a silicon chip integrating SNSPDS directly patterned on to a lithographically defined silicon photonic chip was demonstrated, with application to untrusted MDI QKD nodes (Fig.~\ref{fig:QC}c)~\cite{Zhang19arxivChip}.

Being an indirect bandgap material, silicon lacks accessible light source technolog.
Future application of silicon photonic QKD links will depend on the availability of efficient light sources for the generation of weak coherent pulses on the platform.
Though progress is being made~\cite{zhou2015chip}, today, solutions based on hybrid techniques are prevalent, for example by flip-chip bonded III-IV devices~\cite{song20163d, agnesi2019hong}.
For example, a silicon photon pair source pumped by a silicon photonic with heterogeneously integrated laser~\cite{wang2018photon} has been reported.

\subsection{Future quantum networks}
Though QKD enables the sharing of a secret key between precisely two parties, more general communication scenarios require non-classical states as resources~\cite{markham2008graph, azuma2015all}.
For example, distribution of entangled states enables two quantum states, to be `teleported' between the remote locations.
In this way, it is possible to interconnect multiple and different users around the world, who can utilize non-classical correlation for various applications, from digital transaction and secure-communications to coordination agreements protocols~\cite{kimble2008quantum}.

The first chip-to-chip transportation of quantum states was achieved with two silicon photonic devices via coherent path-to-polarisation state conversion grating coupler (Fig.~\ref{fig:multiph}b) and optical fiber link~\cite{wang2016chip}.
More recently, a four-photon GHZ state was generated on-chip before being partially distributed and being used to teleport a quantum state between discrete chips~\cite{llewellyn2020chip}.
These works extend the overall range of quantum applications using silicon photonics, and light the way for quantum networks based on silicon photonic technology.

Meanwhile, the `holy grail' of limitless quantum communications via quantum repeaters, will be enabled by many of the same technologies required for scaling quantum photonic information processors.
In fact, the mechanics of a quantum repeater---teleporting a qubit from one location to another---are a core feature of measurement-based quantum computation.
All-photonic repeater schemes use graph states~\cite{azuma2015all}---recently demonstrated in silicon photonics~\cite{adcock2019programmable}---in a scheme that has been illustrated in bulk optics~\cite{li2019experimental}.
Eventually, the transmission of quantum data between quantum computers will require quantum interconnects between computer and network qubits.
These will likely require integrated photonics, and will form the basis of an entanglement-based quantum internet~\cite{kimble2008quantum, lee2020quantum}.

\subsection{Quantum sensing}
\label{sec:sensing}

High-fidelity state generation and manipulation also opens the door to quantum sensing applications.
For example, NOON states, which exhibit phase sensitivity beyond the standard quantum, or shot noise, limit ($\propto \sqrt{N}$ samples), and are instead subject to Heisenberg scaling ($\propto N$ samples)~\cite{giovannetti2004quantum, matthews2011heralding}.
Meanwhile, graph states---resources for measurement-based quantum protocols---have also been demonstrated to beat the standard quantum limit, and demonstrate robustness to both dephasing and loss noise~\cite{shettell2020graph}.
A system based on these states of single photons may be useful for applications where too much light might be damage or alter the sample.
Otherwise, brighter states of light may be more useful.

Squeezed states of light are one of the most promising states for quantum sensing~\cite{lawrie2019quantum, lawrie2019quantum, degen2017quantum}, as they are both bright and provide quantum enhanced sensitivitiy in their squeezed quadrature.
Today the foremost application of squeezed states in gravitational waves detectors---the world's most sensitive interferometers~\cite{aasi2013enhanced, tse2019quantum,  mcculler2020frequency}.
Recently, on-chip homodyne detection was demonstrated for the first time by using a silicon photonic chip~\cite{raffaelli2018homodyne}, with immediate application to Gbps generation of quantum random numbers~\cite{raffaelli2018generation}.
This scalable approach could enable the implementation of dozens or hundreds of parallel homodyne detectors in the near future.
Here, close integration of electronic control and readout (on the same die, or directly bonded) will enable increased data rates and detection bandwidths.
Homodyne detectors are also key component of continuous variables quantum information (CVQI) architectures, based on squeezed states~\cite{larsen2019deterministic, asavanant2019generation}, which, like single-photon-based processing, is enabled by adaptive measurements and feed-forward~\cite{takeda2019toward}.
The large state spaces demonstrated in these architectures also relies on a time-multiplexing approach~\cite{larsen2019deterministic, asavanant2019generation}.
Featuring pure sources, high-fidelity optics, and fast switching, the requirements for the two technologies are strikingly similar
Though so far efforts integrating CVQI have focused on silicon nitride~\cite{xanq, vernon2019scalable}, techniques such resonance-selective FWM via dual-ring structures (Fig.~\ref{fig:rings}e) are transferable~\cite{zhang2020single, tan2020stimulated}.
Meanwhile, a nonlinear interferometer was recently demonstrated on a silicon photonic chip~\cite{ono2019observation}, with application to imaging and spectroscopy at a wavelengths with limited detection technology~\cite{lemos2014quantum, kalashnikov2016infrared, zou1991induced}.

\section{Future scaling and outlook}
\label{sec:scaling}

\subsection{Technical challenges in silicon quantum photonics}

Today, silicon quantum photonics is being advanced by research around the world.
It is unclear which techniques and platforms the silicon quantum photonic community will converge to as the field progresses, but for rapid progress, developments in theory and applications must inform fabrication process, and vice versa.
Here, we give an overview of the the main technological hurdles in silicon quantum photonics together with possible routes to progress.

\subsubsection{Truly Gaussian pump pulses} Most theoretical models demonstrating near-unity purtiy assume gaussian pump pulses with deviations known to cause reductions in spectral purity.

\subsubsection{High-resolution lithography} Spectral deviations in lithographically printed sources are typically due to finite fabrication tolerances---improvement will come from increased availability of high-resolution lithography. 

\subsubsection{Ultra-low-loss components \& delay lines} Today's ultra-low-loss components---based on sidewall smoothing techniques---will enable significant near-term scaling, if they are combined with advanced silicon photonics platforms featuring phase modulators and efficient grating couplers. Further ahead, hybrid silica or silicon nitride delay lines could enable the next generation of multiplexed photon sources and feedforward capability
\subsubsection{Fast, low-loss switches} Fast, low-loss switches, together with low-latency electronic control logic form the basis of feed-forward technology and truly scalable quantum photonics, enabling photon multiplexing for deterministic photon sources, as well as adaptive measurement-based quantum protocols. Recent developments in hybrid TFLN and BTO devices are poised to make this breakthrough, with loss reductions continally reaping benefits in scale.
\subsubsection{Electronic-photonic co-integration} With BTO or TFLN switching technology logic latency (and associated loss of delayed photons) will be the bottleneck to scalability. Directly connected (flip-chip or wirebonded) electronic logic will empower the next generation of silicon photonic scaling, with efficient monolithic integration a long term goal.
\subsubsection{Pump rejection} On-chip pump rejection has seen good progress, needed for the integration of SNSPDs. Here, continued work to minimise loss in the single photon channels is important. Advanced processing discussed above, will enable further progress, with more efficient sources with larger channel separation also serving.
\subsubsection{Thin-film superconductor on advanced silicon photonics} Given effective pump-rejection integration of SNSPDs with a low-loss, high-speed silicon photonics platform is a must to reach truly large-scale silicon quantum photonics. 
\subsubsection{Signal multiplexing \& thermal budgeting} Cryogenic integrated photonics (for integrating SNSPDs) comes with severe power dissipation and input/output channel limitations. Low-power phase modulation, signal multiplexing, and cryogenic logic in will be enabling technologies.

\subsection{Outlook}

There is concurrent work on all of these challenges, and, as discussed in the above sections, great progress has been made in the last five years on all of them.
Moving forward, the outstanding challenge is their combination and integration to a single platform.
This will facilitate ever-lower loss and latency, and engender continued increases in complexity, scale---and most importantly---meaningful capability in quantum photonic technology.
How can this be achieved?
A recurrent theme is lithographic process and device fabrication, which are the enabling technologies underpinning silicon quantum photonics.
Complete integration is the goal, with every advance on the road enabling new ground-breaking applications.
Some results---such as those discussed above---we can anticipate eagerly, some breakthroughs are not foreseen here.
With more focus than ever on quantum technology, the coming years are poised to see unprecedented advances in silicon quantum photonics.

\section*{Copyright Notice}

Images in this manuscript are reproduced with permission from the copyright owner indicated in the referenced work, unless otherwise stated.

\section*{Acknowledgments}

We acknowledge support from VILLUM FONDEN, QUANPIC (ref. 00025298), the Center of Excellence, Denmark SPOC (ref DNRF123) and EraNET cofund initiatives QuantERA within the  European Union’s Horizon 2020 research and innovation program grant agreement No.~731473 (project SQUARE).

\bibliographystyle{IEEEtran}
\bibliography{IEEEabrv,Bibliography}

\end{document}